\documentclass[twocolumn,10pt]{IEEEtran}

\usepackage{amsmath}
\usepackage{amsfonts}
\usepackage{epsfig}
\usepackage{amssymb}
\usepackage{cite}
\usepackage{color,soul}
\usepackage{subfigure}
\usepackage{multirow}
\usepackage{rotating}
\usepackage{graphicx}
\usepackage{tabularx}
\usepackage{array}
\usepackage{graphicx,dblfloatfix}
\usepackage{blindtext}
\usepackage{ragged2e}
\usepackage{xcolor,colortbl}
\definecolor{Gray}{gray}{0.85}
\usepackage{amsthm,amssymb,amsmath,bm}
\usepackage{fancyhdr}
\usepackage[subfigure]{tocloft}
\usepackage[font={small}]{caption}
\usepackage{algorithmic}
\usepackage[Algorithm]{algorithm}
\newtheorem*{remark}{Remark}

\begin{document}
 \title{Secure transmission with covert requirement in untrusted relaying networks}
\author{\IEEEauthorblockN{Moslem Forouzesh \textit{Student Member, IEEE}, Paeiz Azmi, \textit{Senior Member, IEEE,} and Ali Kuhestani, \textit{Student Member, IEEE} }  \textsuperscript{}\thanks{\noindent\textsuperscript{} Moslem Forouzesh is with the Department of Electrical and
 		Computer Engineering, Tarbiat Modares University, Tehran, Iran
 		(e-mail: m.Forouzesh@modares.ac.ir).
 		
 		Paeiz Azmi is with the Department of ECE, Tarbiat Modares University,
 		Tehran, Iran (e-mail: pazmi@modares.ac.ir).
 		
 		Ali Kuhestani is with the Department of Electrical
 		Engineering, Amirkabir University of Technology, Tehran, Iran (e-mail: a.kuhestani@aut.ac.ir).}}
 
 \maketitle
  \begin{abstract}
  	In this paper, we study the problem of secure transmission with covert requirement in untrusted relaying networks. Our considered system model consists
  	of one source, one destination, one untrusted relay, and one Willie. The untrusted relay tries to extract the information signal, while the goal of Willie is to detect  the presence of the information signal transmitted by the source, in the current time slot. To overcome these two attacks, we illustrate that the destination and the source should inject jamming signal to the network in phase I and phase II, respectively. Accordingly, the communication in our proposed system model is accomplished in two phases.  In the first
  	phase, when the source transmits its data to the untrusted relay the destination broadcasts its jamming signal. In the second phase, when the relay retransmits the received signal, the source transmits a jamming signal with one of its antennas. For this system model, we propose a power allocation strategy to maximize the instantaneous secrecy rate subject to satisfying the  covert requirements in both of the phases. Since the proposed optimization problem is non-convex, we adopt the Successive Convex Approximation
  	(SCA) approach  to convert it to a convex optimization problem. Next, we extend our system model to a practical system model where there are multiple untrusted relays and multiple Willies under two scenarios of non-colluding Willies and colluding Willies. Our findings highlight that unlike the direct transmission scheme, the achievable secrecy rate of the proposed secure transmission scheme improve as the number of untrusted relays increases.
  	
 \end{abstract}
 
 \section{Introduction}\label{Introduction}
 Privacy and security against eavesdropping play vital role in wireless communication networks and hence, significant attentions should be payed to  these areas.
 To tackle the eavesdropping attack, physical layer security (PLS)  has been widely investigated in different system models since Wyners introduced the wiretap
 channel concept \cite{Wyner}.   Wyner demonstrates when
 the eavesdropper's channel is a degraded version  of the legitimate users’ channel, a positive secrecy rate is achievable,  \cite{Mukherjee}. 
  Toward this end, several techniques have been proposed to enhance the PLS:  transmit beamforming \cite{Khisti}, \cite{A.Khisti}, antenna selection  \cite{Yang}, \cite{Alves}, cooperative techniques   \cite{Zheng}-\cite{m.for}, and artificial noise aided transmission \cite{X.Zhang}-\cite{M.For}.
  
 Relaying is an effective approach for enhancing  energy efficiency, increasing  coverage, and improving throughput  in wireless communications. However, an untrusted relay may
 intentionally overhear the information signal when relaying, i.e., the untrusted relay is a collaborator in service level while it may be an eavesdropper in data level. This scenario occurs in large-scale wireless systems such as heterogeneous networks and  Internet-of-things (IoT) applications, where confidential  messages are often retransmitted by intermediate nodes.
 
 To achieve a positive secrecy rate in untrusted relaying networks, the destination-based cooperative jamming (DBCJ) scheme was first introduced in \cite{X.He}. We note that in the DBCJ the  destination itself contributes to degrade the received signal-to-noise-ratio (SNR) at  the helper node who may act as an eavesdropper. Recently, several  works have taken into account both performance analysis and network optimization of untrusted relaying network \cite{L.Sun}--\cite{Kuhestani_2}.

In some communication networks, low probability of detection (LPD) or covert communication is necessary for data transmission over electromagnetic and acoustic channels \cite{R.Diamant}. For military applications, LPD is interest when the transmitter wishes to remain undetected, or when the knowledge of communication may point to the presence of a receiver. As such, in covert communication only the detection capability of an eavesdropper is considered, i.e., an eavesdropper need not be able to actually decode the communication signal.
 In other words, the  covert communication keeps  military forces from possible attacks \cite{He.B}. In recent years, several papers have investigated the covert communication in different wireless communication networks \cite{jammer}-\cite{Hu}. Specifically, with the idea of employing  a jammer as proposed in \cite{jammer}, source can transmit covertly to Bob in the presence of a careful adversary Willie. The authors in \cite{Forouzesh} studied and compared the performance of the PLS approach and the covert communication approach for a simple wiretap channel with the aim of maximizing the secrecy/covert. Taking into account relaying, the authors in \cite{Hu} examined the achievable performance of covert communication in amplify-and-forward (AF) one-way relay networks. To bes specific, the relay is greedy and opportunistically transmits its own information to the destination covertly besides retransmitting the source’s message, while the source attempts to detect this covert transmission to discover the  illegal usage of the resource dedicated only for the goal of forwarding the source’s information signal.

In this paper, we examine the power allocation problem of an AF  relaying system, where a multiple antenna source transmits its confidential message to a single antenna destination in the presence of a passive eavesdropper. The relay is considered to be both an essential helper and a potential eavesdropper. In our considered network, while the untrusted relay cannot decode the source message, the external eavesdropper is avoided to be aware from the communication in order to provide a strong security in our wireless network. 
 The proposed secure transmission scheme using the untrusted relay is accomplished in two phases. In the first phase, the source transmits data with beamforming technique to the untrusted relay, and simultaneously destination transmits artificial noise, to confuse the curios relay. In the second phase, the relay amplifies the received signal of the first phase, while concurrently the source transmits artificial noise. In the mentioned system model, we take into account the power allocation problem in each phase so that the untrusted relay is not able to extract the information signal and Willie cannot decide about the communication between legitimate parties of the network.
Our key contributions in this paper are summarized as follows:
 \begin{itemize}
\item	
We formulate the power allocation between the source and destination that maximizes the instantaneous secrecy rate of untrusted relaying while concurrently hiding the communication against the passive eavesdropping attack. Since the optimization problem is non-convex, we exploit the Successive Convex Approximation (SCA) approach to convert it to a convex optimization problem.
	\item   
  We extend our system model to a more practical system model where there are multiple untrusted relays and multiple Willies under two scenarios of non-colluding Willies and colluding Willies. For this system model, we develop a simple relay selection criterion. We highlight that the secrecy rate of our proposed untrusted relaying scenario improves with increasing number of relays.
\item
As a benchmark, we investigate the power allocation between the information signal and the jamming signal for the conventional direct transmission dispensing with the relays to compare with the proposed transmission scheme.
\end{itemize}

The remainder of this paper is organized as follows. In Section \ref{System model}, we present system and signal model.   Section \ref{Power Allocation for  Secrecy Rate Maximization With Covert Communication Requirement} provides the detailed problem formulation. In Section \ref{Multi Untrusted Relay Scenario With Relay Selection}, we study multiple untrusted relays and  introduce the relay selection criterion in the considered system model. In Section \ref{Multiple Willies Scenario}, we investigate the scenario with multiple Willies under two cases of non-colluding Willies and colluding Willies. In Section \ref{DT}, direct transmission scheme is studied.  Numerical results are presented in Section \ref{Numerical Results}. Finally, paper is concluded in Section \ref{Conclusion}.
 \section{System and Signal Model}\label{System model}
 
\begin{figure*}[t]  	
	\begin{center}
		\includegraphics[width=7in,height=.8in]{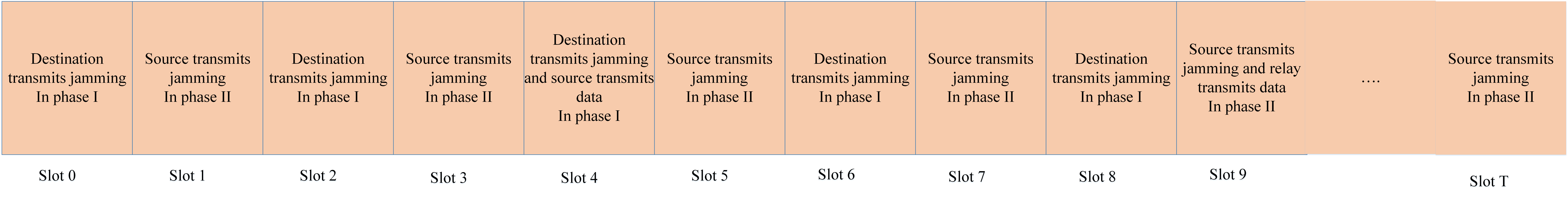}
		\caption{Source decides to transmit data in even slots with probability $pr_t$, also relay decides to transmit data in odd slots  with probability $pr_t$,
			which Willie attempts to detect a transmission in theses slots.}
		\label{TS}
	\end{center}
\end{figure*}

 The system model under investigation is a  one-way relay network consisting of one multiple antenna source with $N_s$ antennas, a single antenna destination, a single antenna untrusted  amplify-and forward (AF) relay, and a single antenna Willie. It should be noted the untrusted relay  is helper  at the service  level while it is untrusted at the data level. The  untrusted relay tries to extract the information signal, while the goal of Willie is to detect  whether the source has sent a signal to destination  or not, in the current time slot. Hence, to increase the error probability of signal detection, the source transmits data in some time slots not during all of the time slots.
 In addition, we assume  all the nodes operate under half-duplex mode. Our proposed network operates at two phases. In the first phase, the source transmits data with beamforming technique toward the untrusted relay, and the destination transmits artificial noise to decrease the signal-to-interference-and-noise ratio (SINR) of the relay and to deceive Willie. In the second phase, the relay normalizes the received signal and broadcasts it. In this time, the source for deceiving Willie emits jamming signal. See Fig \ref{TS}. This communication is performed through a discrete-time channel with $T$  time slots at which each time slot consists of $n$ symbols. The source's data signal and the jamming signal of the source and the destination  in each time slot can be expressed as
 ${\bf{x}}_s^{(1)} = \left[ {x_s^{1(1)},x_s^{2(1)}, \ldots ,x_s^{n(1)}} \right]$,  ${\bf{x}}_s^{(2)} = \left[ {x_s^{1(2)},x_s^{2(2)}, \ldots ,x_s^{n(2)}} \right]$, and ${{\bf{x}}_d} = \left[ {x_d^1,x_d^2, \ldots ,x_d^n} \right]$, respectively.
 
  The complex Gaussian channel between the source and untrusted relay, source and Willie, untrusted relay and Willie, untrusted relay and destination, and destination and Willie are denoted by 
  ${{\bf{h}}_{sr}} \sim {\cal C}{\cal N}\left( {{{\bf{0}}_{{N_s} \times 1}},{\mu _{sr}}{{\bf{I}}_{{N_s} \times 1}}} \right)$, ${{\bf{h}}_{sw}} \sim   {\cal C}{\cal N}\left( {{{\bf{0}}_{{N_s} \times 1}},{\mu _{sw}}{{\bf{I}}_{{N_s} \times 1}}} \right)$, ${h_{rw}} \sim {\cal C}{\cal N}\left( {0,{\mu _{rw}}} \right)$
  , ${h_{rd}}\sim {\cal C}{\cal N}\left( {0,{\mu _{rd}}} \right)$, and ${h_{dw}}\sim {\cal C}{\cal N}\left( {0,{\mu _{dw}}} \right)$, respectively, where $\bf{I}$ is the identity matrix,  $\bf{0}$ is  the zero matrix, 
and   $\mu _{sr}$, $\mu _{sw}$, $\mu _{rw}$, $\mu _{rd}$, $\mu _{dw}$ are variances of each link per branch.
  
 \begin{figure}[t]  
	\begin{center}
		\includegraphics[width=3.8in,height=2.5in]{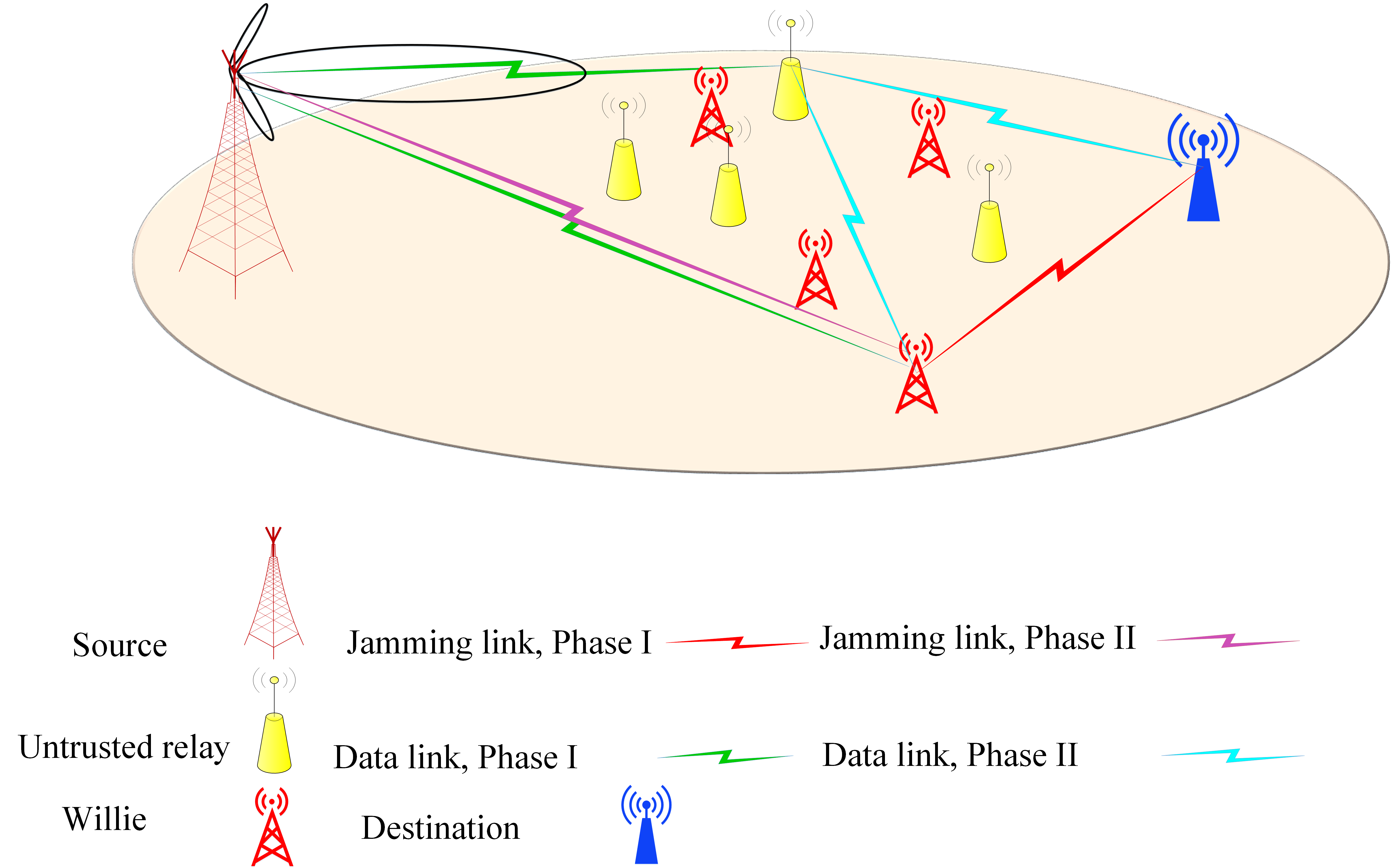}
		\caption{Secure transmission in untrusted relaying network with covert requirement.}
		\label{Sy}
	\end{center}
\end{figure}

The received ${\ell}^{th}$ signal in the each time slot, at untrusted relay in the first phase is given by
\begin{align}
{y_r^{\ell}} =\left\{ {\begin{array}{*{20}{l}}
	{\sqrt {\left( {1 - \rho } \right)P} {h_{rd}}{x_d^{\ell}} + {n_r^{\ell}}},&{{\Psi_0}},\\\\
	{\sqrt {\rho P} {\bf{w}^H}{{\bf{h}}_{sr}}{x_s^{{\ell}(1)}} + \sqrt {\left( {1 - \rho } \right)P} {h_{rd}}{x_d^{\ell}} + {n_r^{\ell}}},&{{\Psi_1}},
	\end{array}} \right.
\end{align}
where $x_s^{{\ell}(1)}$ and $x_d^{\ell}$ are the transmitted ${\ell}^{th}$ symbols  by source and destination, respectively and $n_r^{\ell} \sim {\cal C}{\cal N}\left( {0, \sigma^2} \right) $ is  Additive white Gaussian noise (AWGN) at the untrusted relay. 
Note that,  codebook for $x_d^{\ell}$ is unknown at both relay and Willie  while codebook for $x_s^{{\ell}(1)}$ is known at relay and unknown at Willie.
$P$ is maximum allowable transmit power at each node and $\rho$ is transmit power allocation factor in the first phase.
Since  Willie's CSI is not available, we consider weight vector   ${\bf{w}} = \frac{{{{\bf{h}}_{sr}}}}{{\left\| {{{\bf{h}}_{sr}}} \right\|}}$, which represents the maximum ratio transmission (MRT)  beamformer at the source. 
 Moreover,  notation $\Psi_0$ states that source does not transmit data 
to relay, while $\Psi_1$ states that source transmits data to relay.

 In the second phase, the untrusted relay normalizes the received signal and transmits it with power $\sqrt {\left( {1 - \xi } \right)P } $. In other words, the relay amplifies signal by an amplification factor of 
\begin{align}\label{G}
G_r
=\left\{ {\begin{array}{*{20}{l}}
	{\sqrt {\frac{{\left( {1 - \xi } \right)P}}{{\left( {1 - \rho } \right)P{{\left\| {{h_{rd}}} \right\|}^2} + {\sigma ^2}}}} },&{{\Psi_0}},\\\\
	{\sqrt {\frac{{\left( {1 - \xi } \right) P}}{{\rho P{{\left\| {{{\bf{w}}^H}{{\bf{h}}_{sr}}} \right\|}^2} + \left( {1 - \rho } \right)P{{\left\| {{h_{rd}}} \right\|}^2} + {\sigma ^2}}}} },&{{\Psi_1}},
	\end{array}} \right.
\end{align}
and it transmits ${x_r^{\ell}} = G_r {y_r^{\ell}}$. While the relay broadcasts signal, the source transmits artificial noise to deceive Willie. The received ${\ell}^{th}$ signal at the destination after performing self-interference cancellation is given by
\begin{align}
{y_d^{\ell}}  =\left\{ {\begin{array}{*{20}{l}}
	{ {n_d^{\ell}}},&{{\Psi_0}},\\\\
	{G_r \sqrt {\rho P} {{\bf{w}}^H}{{\bf{h}}_{sr}}{h_{rd}}{x_s^{{\ell}(1)}} + G_r {n_r^{\ell}}{h_{rd}} + {n_d^{\ell}}},&{{\Psi_1}},
	\end{array}} \right.
\end{align}
where $n_d^{\ell} \sim {\cal C}{\cal N}\left( {0, \sigma^2} \right) $ is the AWGN noise at the destination. 

To simplify the mathematical formulation,  we define two parameters $\gamma _{sr}$ and $\gamma _{rd}$ as ${\gamma _{sr}} = \frac{{P{{\left\| {{{\frac{{{\bf{h}_{sr}^H}}}{{\left\| {{{\bf{h}}_{sr}}} \right\|}}}}{{\bf{h}}_{sr}}} \right\|}^2}}}{{{\sigma ^2}}}$ and ${\gamma _{rd}} = \frac{{P{{\left\| {{h_{rd}}} \right\|}^2}}}{{{\sigma ^2}}}$. Hence, 
the received  SINR at the relay and the destination can be written as follows
\begin{align}\label{g_d}
&{\gamma _D} =\left\{ {\begin{array}{*{20}{l}}
	{0},&{{\Psi_0}},\\\\
	{\frac{{{G_r^2}\rho P{{\left\| {{{\frac{{{\bf{h}_{sr}^H}}}{{\left\| {{{\bf{h}}_{sr}}} \right\|}}}}{{\bf{h}}_{sr}}} \right\|}^2}{{\left\| {{h_{rd}}} \right\|}^2}}}{{{G_r^2}{\sigma ^2}{{\left\| {{h_{rd}}} \right\|}^2} + {\sigma ^2}}}\mathop  = \limits^{(a)} \frac{{\rho {\gamma _{sr}}{\gamma _{rd}}\left( {1 - \xi } \right)}}{{ \rho {\gamma _{sr}} + \left( {2 - \rho -\xi} \right){\gamma _{rd}} + 1}}},&{{\Psi_1}},
	\end{array}} \right.
\\&\label{g_r}
{\gamma _R} =\left\{ {\begin{array}{*{20}{l}}
	{0},&{{\Psi_0}},\\\\
	{\frac{{\rho P{{\left\| {{{\frac{{{\bf{h}_{sr}^H}}}{{\left\| {{{\bf{h}}_{sr}}} \right\|}}}}{{\bf{h}}_{sr}}} \right\|}^2}}}{{\left( {1 - \rho } \right)P{{\left\| {{h_{rd}}} \right\|}^2} + {\sigma ^2}}}\mathop  = \limits^{(a)} \frac{{\rho {\gamma _{sr}}}}{{\left( {1 - \rho } \right){\gamma _{rd}} + 1}}},&{{\Psi_1}},
	\end{array}} \right.
\end{align}
where $(a)$  follows from substituting \eqref{G}. Under high  SNR assumption, \eqref{g_d} is simplified as
\begin{align}\label{g_d0}
&{\gamma _D} = \left\{ {\begin{array}{*{20}{l}}
	{0,}&{{\Psi _0},}\\
	{}&{}\\
	{\frac{{\rho {\gamma _{sr}}{\gamma _{rd}}\left( {1 - \xi } \right)}}{{\rho {\gamma _{sr}} + \left( {2 - \rho  - \xi } \right){\gamma _{rd}}}} = \frac{{\rho \varsigma_r {\gamma _{rd}}\left( {1 - \xi } \right)}}{{\rho \varsigma_r  + 2 - \rho  - \xi }},}&{{\Psi _1},}
	\end{array}} \right.
\end{align}
where $\varsigma_r  = \frac{{{\gamma _{sr}}}}{{{\gamma _{rd}}}}$. It is important to notice that deleting $ ``+1"$ from the denominator of $\gamma_R$ is equivalent to  ignore the AWGN noise at the untrusted relay, which is equivalent to the highest  eavesdropping at the relay. Hence, we can rewrite \eqref{g_r} as follow
\begin{align}\label{g_r0}
{\gamma _R} = \left\{ {\begin{array}{*{20}{l}}
	{0,}&{{\Psi _0},}\\
	{}&{}\\
	{\frac{{\rho {\gamma _{sr}}}}{{\left( {1 - \rho } \right){\gamma _{rd}}}} = \frac{{\rho \varsigma_r }}{{1 - \rho }},}&{{\Psi _1},}
	\end{array}} \right.
\end{align}
\section{Power Allocation for  Secrecy Rate Maximization With Covert Communication Requirement}\label{Power Allocation for  Secrecy Rate Maximization With Covert Communication Requirement}
In this section, we aim to maximize the secrecy rate by considering covert requirement at Willie.

\subsection{Physical layer security}\label{Physical layer security}

The instantaneous   secrecy is defined as \cite{Mukherjee}
\begin{align}\label{SR}
{R_{sec}} = \left[ {\log \left( {1 + \gamma_D} \right)} \right.{\left. { - \log \left( {1 + \gamma_R} \right)} \right]^ + },
\end{align}
where $\log \left( {1 + \gamma_D} \right)$ and $\log \left( {1 + \gamma_R} \right)$ are  the achievable information rates at the destination and the untrusted relay, respectively, where ${\left[ v \right]^ + }=\max\left({v, 0}\right)$.

As mentioned before, the source transmits its data during some time slots to hide transmission from Willie. Therefore, by defining $pr_t$ as the probability of signal transmission by the source, the secrecy rate can be rewritten as
\begin{align}\label{secrecy_rate}
R_{sec}=&\frac{pr_t}{2}\left[ {\log \left( {1 + \frac{{\rho \varsigma_r {\gamma _{rd}}\left( {1 - \xi } \right)}}{{\rho \varsigma_r  + 2 - \rho  - \xi }}} \right)} \right.\\&
{\left. { - \log \left( {1 + \frac{{\rho \varsigma_r }}{{1 - \rho }}} \right)} \right]^ + }. \nonumber
\end{align}
Note that for $\rho=0$, the secrecy rate is zero. As such, for optimal power allocation, we have, $R_{sec} \ge 0$ and consequently, \eqref{secrecy_rate} can be rewritten as
\begin{align}
R_{sec}=&\frac{pr_t}{2}\left[ {\log \left( {1 + \frac{{\rho \varsigma_r {\gamma _{rd}}\left( {1 - \xi } \right)}}{{\rho \varsigma_r  + 2 - \rho  - \xi }}} \right)} \right.\\&
{\left. { - \log \left( {1 + \frac{{\rho \varsigma_r }}{{1 - \rho }}} \right)} \right] }. \nonumber
\end{align}
\subsection{Covert requirement}
Based  on the received signal power at Willie, it decides whether the data is transmitted or not. The received ${\ell}^{th}$ signal  at Willie in the first phase is given by
\begin{align}\label{yw1}
&y_w^{{\ell}(1)} =\\& \left\{ {\begin{array}{*{20}{l}}
	{\sqrt {\left( {1 - \rho } \right)P} {h_{dw}}{x_d^{\ell}} + {n_w^{{\ell}(1)}},}&{{\Psi _0},}\\
	{}&{}\\
	{\sqrt {\rho P} {{\bf{w}}^H}{{\bf{h}}_{sw}}{x_s^{\ell(1)}} + \sqrt {\left( {1 - \rho } \right)P} {h_{dw}}{x_d^\ell} + {n_w^{{\ell}(1)}},}&{{\Psi _1},} 
	\end{array}} \right. \nonumber
\end{align}
where $n_w^{{\ell}(1)} \sim {\cal C}{\cal N}\left( {0, \sigma_w^2} \right) $ is  the AWGN noise at Willie. Note that in the covert communication literature, it is assumed that the codebook for data signal $({x_s^{\ell(1)}})$ is unknown at Willie.

 Furthermore, the received ${{\ell}^{th}}$signal  at  Willie in the second phase is given by
\begin{align}\label{yw2}
&{y_w^{{\ell}(2)}} =\\&  \left\{ {\begin{array}{*{20}{l}}
	{\sqrt {\xi P} {h_{sw}}x_s^{{\ell}(2)} + {n_w^{{\ell}(2)}},}&{{\Psi _0},}\\
	{}&{}\\
	\begin{array}{l}
	\left( {\sqrt {\rho P} {{\bf{w}}^H}{{\bf{h}}_{sr}}{x_s^{\ell(1)}} + \sqrt {\left( {1 - \rho } \right)P} {h_{rd}}{x_d^{\ell}} + {n_r}} \right)\times\\G_r{h_{rw}}
	+ \sqrt {\xi P} {h_{sw}}x_s^{{\ell}(2)} + {n_w^{{\ell}(2)}},
	\end{array}&{{\Psi _1},}
	\end{array}} \right. \nonumber
\end{align}
where $n_w^{{\ell}(2)} \sim {\cal C}{\cal N}\left( {0, \sigma_w^2} \right) $ is  the AWGN noise at Willie in second phase.
In the covert communication, Willie selects a threshold and based on comparison between total received power and selected threshold, decides whether source transmits data  or not. In other words,  Willie decides in the first and second phase   based on
$\frac{{{Y_w^{(1)}}}}{n}	\mathop \gtrless\limits_{\Psi_0}^{\Psi_1} \vartheta^{(1)},$ and $\frac{{{Y_w^{(1)}}}}{n}	\mathop \gtrless\limits_{\Psi_0}^{\Psi_1} \vartheta^{(2)},$ respectively, which is demonstrated in \cite{jammer}, this rule is optimal decision rule in points of view Willie.
where ${Y_w} = {\sum\limits_{\ell  = 1}^n {\left| {y_w^\ell } \right|} ^2}$
is defined as the total power received by Willie during a time slot.

 When the source transmits data but Willie decides $\Psi_0$, miss detection with probability of  ${\mathbb{P}_{MD}} $ happens, while if   source does not transmit data but Willie decides $\Psi_1$, false alarm with probability of  ${\mathbb{P}_{FA}} $ happens. While the following constraint is satisfied:
 \begin{align}\label{decision}
 \text{for any}\, \varepsilon \ge 0, \,\,\, \mathbb{P}_{MD}+\mathbb{P}_{FA}\ge 1-\varepsilon, \,\,\, \text{as}\,\,\, n \to \infty.
 \end{align}
 we can say which our communication is covert, \cite{jammer}. The false alarm and miss detection probabilities in the first phase are respectively, given by
\begin{align}\label{P1}
&{\mathbb{P}^{(1)}_{FA}} = \\&\mathbb{P}\left( {\frac{{{Y^{(1)}_w}}}{n} > \vartheta^{(1)} \left| {{\Psi_0}} \right.} \right) = \mathbb{P}\left( {\left( {\sigma _w^2 + \gamma^{(1)} } \right)\frac{{\chi _{2n}^2}}{n} > \vartheta^{(1)} \left| {{\Psi_0}} \right.} \right), \nonumber\\&
{\mathbb{P}^{(1)}_{MD}} = \\& \mathbb{P}\left( {\frac{{{Y^{(1)}_w}}}{n} < \vartheta^{(1)} \left| {{\Psi_1}} \right.} \right) = \mathbb{P}\left( {\left( {\sigma _w^2 + \gamma^{(1)} } \right)\nonumber\frac{{\chi _{2n}^2}}{n} < \vartheta^{(1)} \left| {{\Psi_1}} \right.} \right),
\end{align}
where ${\chi _{2n}^2}$ presents a chi-squared random variable, with $2n$ degrees of freedom.
Furthermore by using \eqref{yw1}$, \gamma ^{(1)}$ is expressed as
\begin{align}\label{gamma1}
{\gamma ^{(1)}} = \left\{ {\begin{array}{*{20}{l}}
	{\left( {1 - \rho } \right)P{{\left| {{h_{dw}}} \right|}^2},}&{{\Psi _0},}\\
	{}&{}\\
	{\rho P{{\left\| {\frac{{{\bf{h}}_{{\bf{sr}}}^{{H}}}}{{\left\| {{{\bf{h}}_{sr}}} \right\|}}{{\bf{h}}_{sw}}} \right\|}^2} + \left( {1 - \rho } \right)P{{\left| {{h_{dw}}} \right|}^2},}&{{\Psi _1}.}
	\end{array}} \right.
\end{align}
Moreover, based on same argument, the false alarm and miss detection probabilities in second phase are respectively, defined as
\begin{align}\label{p2}
&{\mathbb{P}^{(2)}_{FA}} =\\& \mathbb{P}\left( {\frac{{{Y^{(2)}_w}}}{n} > \vartheta^{(2)} \left| {{\Psi_0}} \right.} \right) = \mathbb{P}\left( {\left( {\sigma _w^2 + \gamma^{(2)} } \right)\frac{{\chi _{2n}^2}}{n} > \vartheta^{(2)} \left| {{\Psi_0}} \right.} \right), \nonumber\\&
{\mathbb{P}^{(2)}_{MD}} =\\& \mathbb{P}\left( {\frac{{{Y^{(2)}_w}}}{n} < \vartheta^{(2)} \left| {{\Psi_1}} \right.} \right) = \mathbb{P}\left( {\left( {\sigma _w^2 + \gamma^{(2)} } \right)\frac{{\chi _{2n}^2}}{n} < \vartheta^{(2)} \left| {{\Psi_1}} \right.} \right).\nonumber
\end{align}
By employing \eqref{yw2}$, \gamma ^{(2)}$ can be written as follows
\begin{align}\label{gamma2}
{\gamma ^{(2)}} = \left\{ {\begin{array}{*{20}{l}}
	{\xi P{{\left| {{h_{sw}}} \right|}^2},}&{{\Psi _0},}\\
	{}&{}\\
	{\left( {1 - \xi } \right)P{{\left| {{h_{rw}}} \right|}^2} + \xi P{{\left| {{h_{sw}}} \right|}^2},}&{{\Psi _1},}
	\end{array}} \right.
\end{align} 
According to the strong law of large numbers (SLLN),
$\frac{\chi _{2n}^2}{n}$  converges to 1, and based on 
lebesgue’s dominated convergence theorem \cite{coverage}
we can replace  $\frac{\chi _{2n}^2}{n}$  with 1, when $n \to \infty $. Hence, the false alarm and the miss detection probabilities in the first phase can be rewritten as follows
\begin{align}\label{P11}
{\mathbb{P}^{(1)}_{FA}} = \mathbb{P}\left( {\sigma _w^2 + \gamma^{(1)} > \vartheta^{(1)} \left| {{\Psi_0}} \right.} \right), \\
{\mathbb{P}^{(1)}_{MD}} = \mathbb{P}\left( {\sigma _w^2 + \gamma^{(1)}< \vartheta^{(1)} \left| {{\Psi_1}} \right.} \right),
\end{align}
also in the second phase, we haves:
\begin{align}
{\mathbb{P}^{(2)}_{FA}} = \mathbb{P}\left( {\sigma _w^2 + \gamma^{(2)} > \vartheta^{(2)} \left| {{\Psi_0}} \right.} \right), \\\label{P21}
{\mathbb{P}^{(2)}_{MD}} = \mathbb{P}\left( {\sigma _w^2 + \gamma^{(2)}< \vartheta^{(2)} \left| {{\Psi_1}} \right.} \right).
\end{align}
For solving \eqref{P11}-\eqref{P21}, we need to  probability density function (PDF) of $\gamma ^{(1)}$ and $\gamma ^{(2)}$, hence, we approximate one term in $\gamma ^{(1)}$.
By exploiting the law of large numbers for large $N_s$ \cite{Larsson,Rawat}, we have the following approximation
\begin{align}\label{P_21}
{\left\| {\frac{{{\bf{h}}_{{\bf{sr}}}^{{H}}}}{{\left\| {{{\bf{h}}_{sr}}} \right\|}}{{\bf{h}}_{sw}}} \right\|^2} = \frac{{{{\left\| {{\bf{h}}_{sr}^{{H}}{{\bf{h}}_{sw}}} \right\|}^2}}}{{{{\left\| {{{\bf{h}}_{sr}}} \right\|}^2}}} \approx \Lambda= \frac{{{{\left\| {{\bf{h}}_{sr}^{{H}}{{\bf{h}}_{sw}}} \right\|}^2}}}{{{N_s}{\mu _{sr}}}}
\end{align}
moreover, by utilization of Lindeberg-Levy central limit theorem,
distribution of  random variable ${{\bf{h}}_{sr}^{{H}}{{\bf{h}}_{sw}}}$  can be approximated as $\mathcal{CN}\left( {0,{N_s}{\mu _{sr}}{\mu _{sw}}} \right)$, hence, we can consider $\Lambda$
as an exponential random variable 
with mean $\mu_{sw}$. 
Finally, we  have the following PDFs for  $\gamma ^{(1)}$ and $\gamma ^{(2)}$, respectively,
\begin{align}\label{pdfg1}
&{f_{{\Gamma _1}}}\left( {{\gamma ^{(1)}}} \right) =\nonumber\\& \left\{ {\begin{array}{*{20}{l}}
	{\frac{1}{{\left( {1 - \rho } \right)P\mu_{dw}}}{e^{ - \frac{{{\gamma ^{(1)}}}}{{\left( {1 - \rho } \right)P\mu_{dw}}}}},}&{{\Psi _0},}\\
	{}&{}\\
	{\frac{1}{{P\left( {\rho {\mu _{sw}} - \left( {1 - \rho } \right){\mu _{dw}}} \right)}}\left[ {{e^{ - \frac{{{\gamma ^{(1)}}}}{{\rho P{\mu _{sw}}}}}} - {e^{ - \frac{{{\gamma ^{(1)}}}}{{\left( {1 - \rho } \right)P{\mu _{dw}}}}}}} \right],}&{{\Psi _1}.}
	\end{array}} \right.
\end{align}
\begin{align}\label{pdfg2}
&{f_{{\Gamma _2}}}\left( {{\gamma ^{(2)}}} \right) = \nonumber \\&\left\{ {\begin{array}{*{20}{l}}
	{\frac{1}{{\xi P \mu_{sw}}}{e^{ - \frac{{{\gamma ^{(2)}}}}{\xi P\mu_{sw}}}},}&{{\Psi _0}},\\
	{}&{}\\
	{\frac{1}{{P\left( {\xi {\mu _{sw}} - \left( {1 - \xi } \right){\mu _{rw}}} \right)}}\left[ {{e^{ - \frac{{{\gamma ^{(2)}}}}{{\xi P{\mu _{sw}}}}}} - {e^{ - \frac{{{\gamma ^{(2)}}}}{{\left( {1 - \xi } \right)P{\mu _{rw}}}}}}} \right],}&{{\Psi _1}.}
	\end{array}} \right.
\end{align} 
By employing \eqref{pdfg1}, \eqref{pdfg2} and substituting them into  \eqref{P11}-\eqref{P21}, and  after some mathematical manipulation we have
\begin{align}
\label{fa1}&{\mathbb{P}^{(1)}_{FA}} = \left\{ {\begin{array}{*{20}{l}}
	{{e^{ - \frac{1}{{\left( {1 - \rho } \right)P{\mu _{dw}}}}\left( {{\vartheta ^{(1)}} - {\sigma_w ^2}} \right)}},}&{{\vartheta ^{(1)}} - {\sigma _w^2} > 0,}\\
	{}&{}\\
	{1,}&{{\vartheta ^{(1)}} - {\sigma _w^2} \le 0.}
	\end{array}} \right.
\\
&\label{md1}{\mathbb{P}^{(1)}_{MD}} =\\& \left\{ {\begin{array}{*{20}{l}}
	{\begin{array}{*{20}{l}}
		\begin{array}{l}
		1 + \frac{1}{{\rho {\mu _{sw}} - \left( {1 - \rho } \right){\mu _{dw}}}}\\
		\times \left[ { - \rho {\mu _{sw}}{e^{ - \frac{1}{{\rho P{\mu _{sw}}}}\left( {{\vartheta ^{(1)}} - {\sigma _w^2}} \right)}}} \right.
		\end{array}\\
		{\left. { + \left( {1 - \rho } \right){\mu _{dw}}{e^{ - \frac{1}{{\left( {1 - \rho } \right)P{\mu _{dw}}}}\left( {{\vartheta ^{(1)}} - {\sigma _w^2}} \right)}}} \right]}
		\end{array}}&{{\vartheta ^{(2)}} - {\sigma _w^2} > 0,}\\
	{\begin{array}{*{20}{l}}
		{}
		\end{array}}&{}\\
	0&{{\vartheta ^{(2)}} - {\sigma_w ^2} \le 0.}
	\end{array}} \right. \nonumber
\end{align}
\begin{align}
\label{fa2}&{\mathbb{P}^{(2)}_{FA}}  = \left\{ {\begin{array}{*{20}{l}}
	{{e^{ - \frac{1}{{\xi P{\mu _{sw}}}}\left( {{\vartheta ^{(2)}} - {\sigma _w^2}} \right)}},}&{{\vartheta ^{(2)}} - {\sigma _w^2} > 0,}\\
	{}&{}\\
	{1,}&{{\vartheta ^{(2)}} - {\sigma _w^2} \le 0.}
	\end{array}} \right.
\\&
\label{md2}{\mathbb{P}^{(2)}_{MD}}  = \\& \left\{ {\begin{array}{*{20}{l}}
	{\begin{array}{*{20}{l}}
		\begin{array}{l}
		1 + \frac{1}{{\xi {\mu _{sw}} - \left( {1 - \xi } \right){\mu _{rw}}}}\\
		\times \left[ { - \xi {\mu _{sw}}{e^{ - \frac{1}{{\xi P{\mu _{sw}}}}\left( {{\vartheta ^{(2)}} - {\sigma _w^2}} \right)}}} \right.
		\end{array}\\
		{\left. { + \left( {1 - \xi } \right){\mu _{rw}}{e^{ - \frac{1}{{\left( {1 - \xi } \right)P{\mu _{rw}}}}\left( {{\vartheta ^{(2)}} - {\sigma _w^2}} \right)}}} \right],}
		\end{array}}&{{\vartheta ^{(2)}} - {\sigma _w^2} > 0,}\\
	{\begin{array}{*{20}{l}}
		{}
		\end{array}}&{}\\
	0&{{\vartheta ^{(2)}} - {\sigma _w^2} \le 0.}
	\end{array}} \right. \nonumber
\end{align}
According to \eqref{fa1}-\eqref{md2}, Willie try to selects decision threshold  greater than variance of noise in both phases, in other words it selects $\vartheta ^{(1)}>\sigma_w^2$ and $\vartheta ^{(2)}>\sigma_w^2$.  Because in otherwise, $\mathbb{P}^{(1)}_{FA}+\mathbb{P}^{(1)}_{MD}=1$ and  $\mathbb{P}^{(2)}_{FA}+\mathbb{P}^{(2)}_{MD}=1$.
\begin{remark} [The requirement for jamming signal transmission in each phase]
 We illustrate, if destination and source do not transmit jamming signal in phase 1 and 2, respectively, the minimum error probability is near to zero, i.e., covert requirement is not satisfied. Such as, in phase 1, without transmission jamming signal, we have
  \begin{align}
 \begin{array}{*{20}{l}}
 {}&{\mathbb{P}_{FA}^{(1)} = \left\{ {\begin{array}{*{20}{l}}
 		{0,}&{{\vartheta ^{(1)}} - {\sigma _w^2} > 0,}\\
 		{}&{}\\
 		{1,}&{{\vartheta ^{(1)}} - {\sigma _w^2} \le 0.}
 		\end{array}} \right.}\\
 {}&{\mathbb{P}_{MD}^{(1)} = }\\
 {}&{\left\{ {\begin{array}{*{20}{l}}
 		{\begin{array}{*{20}{l}}
 			{1 - {{e^{ - \frac{1}{{\rho P{\mu _{sw}}}}\left( {{\vartheta ^{(1)}} - {\sigma _w^2}} \right)}}}}\\
 			{}
 			\end{array}}&{{\vartheta ^{(1)}} - {\sigma _w^2} > 0,}\\
 		{\begin{array}{*{20}{l}}
 			{}
 			\end{array}}&{}\\
 		0&{{\vartheta ^{(1)}} - {\sigma_w ^2} \le 0.}
 		\end{array}} \right.{\rm{ }}}
 \end{array}
 \end{align}
 It is clear that $\mathop {\min}\limits_{\vartheta^{(1)}}\left(\;{\mathbb{P}^{(1)}_{FA}} + {\mathbb{P}^{(1)}_{MD}}\right)\simeq 0$. Likewise,  $\mathop {\min}\limits_{\vartheta^{(2)}}\left(\;{\mathbb{P}^{(2)}_{FA}} + {\mathbb{P}^{(2)}_{MD}}\right)\simeq 0$. 
\end{remark}
\subsection{power allocation optimization problem} \label{pa_subsection}
In this section, we formulate an optimization problem, with the goal of maximizing the achievable secrecy rate by considering the covert requirement at Willie, i.e., \eqref{decision} is satisfied. Hence, the optimization problem is formulated as
 \begin{subequations}\label{OPT}
	\begin{align}
	&\max_{\rho, \xi}\; \frac{pr_t}{2}\left[ {\log \left( {1 + \frac{{\rho \varsigma_r {\gamma _{rd}}\left( {1 - \xi } \right)}}{{\rho \varsigma_r  + 2 - \rho  - \xi }}} \right)} \right.\\&
	{\left. { - \log \left( {1 + \frac{{\rho \varsigma_r }}{{1 - \rho }}} \right)} \right] }, \nonumber
	\\& \label{rho_limited}
	\hspace{-.5cm}\text{s.t.}:\hspace{.18cm}  \hspace{.008cm} 0 \le \rho\le 1, 
	\\& \hspace{.35cm}\hspace{.008cm}‌0 \le \xi\le 1, \label{xi_limited} 
	\\& \hspace{.35cm} \min_{\vartheta^{(1)}}\left(\;{\mathbb{P}^{(1)}_{FA}} + {\mathbb{P}^{(1)}_{MD}}\right)\ge 1-\varepsilon \label{min_ph1},
	\\& \hspace{.35cm}  \min_{\vartheta^{(2)}}\left(\;{\mathbb{P}^{(2)}_{FA}} + {\mathbb{P}^{(2)}_{MD}}\right)\ge 1-\varepsilon \label{min_ph2},
	\end{align}
 \end{subequations}
Constraints \eqref{rho_limited} and \eqref{xi_limited} specify the range of power allocation factors in the first and second phases, respectively. Constraints \eqref{min_ph1} and \eqref{min_ph2} present the worst-case covert rate requirement at Wilie in first and second phases, respectively.  In order to solve \eqref{OPT}, we first solve \eqref{min_ph1} and \eqref{min_ph2}, to obtain optimal $\vartheta^{(1)}$ and $\vartheta^{(2)}$ from Wilie perspective.
\subsubsection{Optimal Threshold for Willie}
For obtaining optimal  $\vartheta^{(1)}$ and $\vartheta^{(2)}$, we derivative from ${\mathbb{P}^{(1)}_{FA}} + {\mathbb{P}^{(1)}_{MD}}$ and ${\mathbb{P}^{(2)}_{FA}} + {\mathbb{P}^{(2)}_{MD}}$ respect to $\vartheta^{(1)}$ and $\vartheta^{(2)}$, respectively. After calculation 
$\frac{{d\left( {\mathbb{P}^{(1)}_{FA}} + {\mathbb{P}^{(1)}_{MD}} \right)}}{{d\vartheta^{(1)}}}=0$ and $\frac{{d\left( {\mathbb{P}^{(2)}_{FA}} + {\mathbb{P}^{(2)}_{MD}} \right)}}{{d\vartheta^{(2)}}}=0$
and some mathematical manipulations, we have
\begin{align}
\label{thete_opt1} &\vartheta^{(1)}_{opt}=\frac{{{\lambda^{(1)} _j}{\lambda^{(1)} _s}}}{{{\lambda^{(1)} _s} - {\lambda^{(1)} _j}}}\ln \left( {\frac{{{\lambda^{(1)} _s}}}{{{\lambda^{(1)} _j}}}} \right) +\sigma _w^2, \\&\label{thete_opt2} \vartheta^{(2)}_{opt}=\frac{{{\lambda^{(2)} _j}{\lambda^{(2)} _s}}}{{{\lambda^{(2)} _s} - {\lambda^{(2)} _j}}}\ln \left( {\frac{{{\lambda^{(2)} _s}}}{{{\lambda^{(2)} _j}}}} \right)+\sigma _w^2,
\end{align}
where 
$
\lambda^{(1)} _j={\left( {1 - \rho } \right)P{\mu _{dw}}}$, $\lambda^{(1)} _s=\rho P{\mu _{sw}} $, $\lambda^{(2)} _j=\xi P{\mu _{sw}}$, and $\lambda^{(2)} _s=\left( {1 - \xi } \right)P{\mu _{rw}}.
$
By substituting \eqref{thete_opt1} and \eqref{thete_opt2} into \eqref{fa1}, \eqref{md1}, \eqref{fa2}, and \eqref{md2} we have 

\begin{align}\label{sum_fa_md_1}
&\mathbb{P}_{FA}^{(1)} + \mathbb{P}_{MD}^{(1)} ={e^{ - \frac{{\lambda _s^{(1)}}}{{\lambda _s^{(1)} - \lambda _j^{(1)}}}\ln \left( {\frac{{\lambda _s^{(1)}}}{{\lambda _j^{(1)}}}} \right)}} + \frac{1}{{\lambda _s^{(1)} - \lambda _j^{(1)}}} \times \\& \left[ { - \lambda _s^{(1)}{e^{ - \frac{{\lambda _j^{(1)}}}{{\lambda _s^{(1)} - \lambda _j^{(1)}}}\ln \left( {\frac{{\lambda _s^{(1)}}}{{\lambda _j^{(1)}}}} \right)}} + \lambda _j^{(1)}{e^{ - \frac{{\lambda _s^{(1)}}}{{\lambda _s^{(1)} - \lambda _j^{(1)}}}\ln \left( {\frac{{\lambda _s^{(1)}}}{{\lambda _j^{(1)}}}} \right)}}} \right] + 1, \nonumber
\end{align}
\begin{align}\label{sum_fa_md_2}
&\mathbb{P}_{FA}^{(2)} +\mathbb{P} _{MD}^{(2)} = {e^{ - \frac{{\lambda _s^{(2)}}}{{\lambda _s^{(2)} - \lambda _j^{(2)}}}\ln \left( {\frac{{\lambda _s^{(2)}}}{{\lambda _j^{(2)}}}} \right)}} + \frac{1}{{\lambda _j^{(2)} - \lambda _s^{(2)}}} \times \\& \left[ { - \lambda _j^{(2)}{e^{ - \frac{{\lambda _s^{(2)}}}{{\lambda _s^{(2)} - \lambda _j^{(2)}}}\ln \left( {\frac{{\lambda _s^{(2)}}}{{\lambda _j^{(2)}}}} \right)}} + \lambda _s^{(2)}{e^{ - \frac{{\lambda _j^{(2)}}}{{\lambda _s^{(2)} - \lambda _j^{(2)}}}\ln \left( {\frac{{\lambda _s^{(2)}}}{{\lambda _j^{(2)}}}} \right)}}} \right] + 1. \nonumber
\end{align}
By employing \eqref{sum_fa_md_1} and \eqref{sum_fa_md_2} the  optimization problem \eqref{OPT} can be reformulated as

\begin{subequations}\label{OPT_reformulated1}
	\begin{align}
	&\max_{\rho, \xi}\; \frac{pr_t}{2}\left[ {\log \left( {1 + \frac{{\rho \varsigma_r {\gamma _{rd}}\left( {1 - \xi } \right)}}{{\rho \varsigma_r  + 2 - \rho  - \xi }}} \right)} \right.\\&
	{\left. { - \log \left( {1 + \frac{{\rho \varsigma_r }}{{1 - \rho }}} \right)} \right] }, \nonumber
	\\&
	\hspace{-.5cm}\text{s.t.}:\hspace{.18cm} \hspace{.008cm} \eqref{rho_limited}, \eqref{xi_limited},
	\\& \hspace{.35cm}  \ln \left( {\frac{{\lambda _j^{(1)}}}{{\lambda _s^{(1)}}}} \right)\left( {\frac{{\lambda _j^{(1)}}}{{\lambda _s^{(1)} - \lambda _j^{(1)}}}} \right) \le \ln \left( \varepsilon  \right) \label{min_ph},
	\\& \hspace{.35cm}  \ln \left( {\frac{{\lambda _j^{(2)}}}{{\lambda _s^{(2)}}}} \right)\left( {\frac{{\lambda _j^{(2)}}}{{\lambda _s^{(2)} - \lambda _j^{(2)}}}} \right) \le \ln \left( \varepsilon  \right) \label{min_ph0}.
	\end{align}
\end{subequations}
Finally,  we should solve the following optimization problem:
\begin{subequations}\label{OPT_reformulated2}
	\begin{align}
	&\max_{\rho, \xi}\; \frac{pr_t}{2}\left[ {\log \left( {1 + \frac{{\rho \varsigma_r {\gamma _{rd}}\left( {1 - \xi } \right)}}{{\rho \varsigma_r  + 2 - \rho  - \xi }}} \right)} \right.\\&
	{\left. { - \log \left( {1 + \frac{{\rho \varsigma_r }}{{1 - \rho }}} \right)} \right] }, \nonumber
	\\&
	\hspace{-.5cm}\text{s.t.}:\hspace{.18cm} \hspace{.008cm} \eqref{rho_limited}, \eqref{xi_limited},
	\\&‌\hspace{.35cm} \ln \left( {\frac{{\left( {1 - \rho } \right){\mu _{dw}}}}{{\rho {\mu _{sw}}}}} \right)\left( {\frac{{\left( {1 - \rho } \right){\mu _{dw}}}}{{\rho {\mu _{sw}} - \left( {1 - \rho } \right){\mu _{dw}}}}} \right) \le \ln \left( \varepsilon  \right)
	\\& \hspace{.35cm}  \ln \left( {\frac{{\xi {\mu _{sw}}}}{{\left( {1 - \xi } \right){\mu _{rw}}}}} \right)\left( {\frac{{\xi {\mu _{sw}}}}{{\left( {1 - \xi } \right){\mu _{rw}} - \xi {\mu _{sw}}}}} \right) \le \ln \left( \varepsilon  \right),
	\end{align}
\end{subequations}
By utilization the axillary variables $t_0$ and $t_1$ and high SINR assumption,  the optimization problem \eqref{OPT_reformulated2} is equivalent to the following problem

\begin{subequations}\label{OPT_reformulated3}
	\begin{align}
	&\max_{\rho, \xi, t_0, t_1}\; \frac{pr_t}{2}\left[ {\log \left( { \frac{{\rho \varsigma_r {\gamma _{rd}}\left( {1 - \xi } \right)}}{{\rho \varsigma_r  + 2 - \rho  - \xi }}} \right)} \right.\\&
	{\left. { - \log \left( {1 + \frac{{\rho \varsigma_r }}{{1 - \rho }}} \right)} \right] }, \nonumber
	\\&
	\hspace{-.5cm}\text{s.t.}:\hspace{.18cm} \hspace{.008cm}\eqref{rho_limited}, \eqref{xi_limited},
	\\&‌\hspace{.35cm} \ln \left( {\frac{{\left( {1 - \rho } \right){\mu _{dw}}}}{{\rho {\mu _{sw}}}}} \right) \times \left( {1 - \rho } \right){\mu _{dw}} - {t_0}\ln \left( \varepsilon  \right) \le 0, \label{cp01}
	\\&‌\hspace{.35cm} \rho {\mu _{sw}} - \left( {1 - \rho } \right){\mu _{dw}} \le {t_0}, \label{cp11}
	\\& \hspace{.35cm}  \ln \left( {\frac{{\xi {\mu _{sw}}}}{{\left( {1 - \xi } \right){\mu _{rw}}}}} \right) \times \xi {\mu _{sw}} - {t_1}\ln \left( \varepsilon  \right) \le 0, \label{cp02}
	\\& \hspace{.35cm} \left( {1 - \xi } \right){\mu _{rw}} - \xi {\mu _{sw}} \le {t_1}. \label{cp12}
	\end{align}
\end{subequations}
This optimization problem is non-convex yet, because the objective function is non-concave. To tackle this problem, we employ Successive Convex approximation (SCA) approach to approximate the objective function to concave function.   The objective function  can be rewritten as follows
\begin{align}\label{obj_func_diff}
&\log \left( {\rho {\varsigma _r}{\gamma _{rd}}\left( {1 - \xi } \right)} \right) + \log \left( {1 - \rho } \right) - \log \left( {\rho {\varsigma _r} + 2 - \rho  - \xi } \right) - \nonumber\\& \log \left( {1 - \rho  + \rho {\varsigma _r}} \right).
\end{align}
Since \eqref{obj_func_diff} is the difference between two concave functions, we can employ Difference of two Concave functions (DC) method to convert it to the concave function. Hence, we can write \eqref{obj_func_diff} as $ \Xi \left( {\rho ,\xi } \right) = \Sigma  \left( {\rho ,\xi } \right) - \Omega \left( {\rho ,\xi } \right)$, where $\Sigma  \left( {\rho ,\xi } \right)=\log \left( {\rho {\varsigma _r}{\gamma _{rd}}\left( {1 - \xi } \right)} \right) + \log \left( {1 - \rho } \right)$ and $\Omega \left( {\rho ,\xi } \right)=\log \left( {\rho {\varsigma _r} + 2 - \rho  - \xi } \right)+ \log \left( {1 - \rho  + \rho {\varsigma _r}} \right)$. By employing DC method, we can rewrite $\Omega \left( {\rho ,\xi } \right)$ as follows
\begin{align}
&\Omega \left( {\rho ,\xi } \right) \simeq \tilde \Omega \left( {\rho ,\xi } \right) = \Omega \left( {\rho \left( {\mu  - 1} \right),\xi \left( {\mu  - 1} \right)} \right) +  \\&
{\nabla ^T}\Omega \left( {\rho \left( {\mu  - 1} \right),\xi \left( {\mu  - 1} \right)} \right)  \cdot  \left[ {\rho  - \rho \left( {\mu  - 1} \right),\xi  - \xi \left( {\mu  - 1} \right)} \right],\nonumber
\end{align}
where $\mu$ is iteration number and ${\nabla}\Omega \left( {\rho ,\xi} \right)$ is gradient of $\Omega \left( {\rho ,\xi} \right)$ which is calculated as
\begin{align}
&\nabla \Omega \left( {\rho ,\xi } \right) = \nonumber \\& \left[ {\frac{{{\varsigma _r} - 1}}{{\rho {\varsigma _r} + 2 - \rho  - \xi }} + \frac{{{\varsigma _r} - 1}}{{1 - \rho  + \rho {\varsigma _r}}},\frac{{ - 1}}{{\rho {\varsigma _r} + 2 - \rho  - \xi }}} \right].
\end{align}
Finally, by employing DC, we face to a convex optimization problem as follows  :
	\begin{align}\label{Convex_p}
	&\max_{\rho, \xi, t_0, t_1}\; \frac{pr_t}{2} \left( {\Sigma \left( {\rho ,\xi } \right) - \tilde \Omega \left( {\rho ,\xi } \right)} \right)
	\\&
	\hspace{-.5cm}\text{s.t.}:\hspace{.18cm} \hspace{.008cm}\eqref{rho_limited}, \eqref{xi_limited}, \eqref{cp01}, \eqref{cp11}, \eqref{cp02},  \eqref{cp12}, \nonumber
	\end{align}
We employ available softwares, such as CVX solver to solve convex optimization problem \eqref{Convex_p}. Moreover, as we use DC method, we should employ iterative algorithm as Algorithm \ref{alg1}.
\begin{algorithm}[t]
	\caption{ITERATIVE POWER ALLOCATION ALGORITHM} \label{alg1}
	\begin{algorithmic}[1]
		\STATE  \nonumber
		Initialization: Set $\mu =0
		\left( {\mu \text{\hspace{.2cm}is the iteration number}} \right)$
		and initialize to $\rho(0)$ and $\xi(0)$.
		\STATE \label{set}
		Set $\rho=\rho(\mu)$ and $\xi=\xi(\mu)$,
		\STATE  		
		Solve \eqref{Convex_p} and set  the result  to $\rho(\mu +1)$ and $\xi(\mu +1)$
		\STATE 
		If $\left| {\rho\left( {\mu  + 1} \right) - \rho \left( \mu  \right)} \right| \le \theta  $ and $\left| {\xi\left( {\mu  + 1} \right) - \xi \left( \mu  \right)} \right| \le \theta  $\\
		stop,\\
		else\\
		set $\mu = \mu + 1$ and go back to step \ref{set}
	\end{algorithmic}
\end{algorithm}
This algorithm is ended when the stopping conditions i.e.,  $\left| {\rho\left( {\mu  + 1} \right) - \rho \left( \mu  \right)} \right| \le \theta  $ and $\left| {\xi\left( {\mu  + 1} \right) - \xi \left( \mu  \right)} \right| \le \theta  $ are satisfied, where $\theta$ is stopping threshold. 

\section{Multi Untrusted Relay Scenario With Relay Selection}\label{Multi Untrusted Relay Scenario With Relay Selection}
In this section, we develop our system model to multi untrusted relay when the relays adopt  selection combining (SC) technique. In the first phase, the source transmits the message signal to a selected relay using the MRT beamformer. It is clear in this phase,  all the relays  intercept the transmissions and then attempt to extract the confidential information. Moreover, in this phase, the destination transmits jamming signal similar to prior scenario. In the second phase, the selected relay broadcasts data and hence, all the non-selected relays eavesdrop the emitted signal by the selected relay. In this phase, we propose the source transmits jamming signal to confuse both Willie and the other relays.
\subsection{Signal Model in Relay Selection Scenario }

The received signal at relay $\jmath$ ($\jmath  \in \left\{ {1,2,...J} \right\}$) where $J$ is the total numbers of relay, can be written as
\begin{align}\label{yr1_s2}
&{y_\jmath^{\ell(1)}} \nonumber=\\&\left\{ {\begin{array}{*{20}{l}}
	{\sqrt {\left( {1 - \rho } \right)P} {h_{\jmath d}}{x_d^{\ell}} + {n_\jmath^{\ell(1)}}},&{{\Psi_0}},\\\\
	{\sqrt {\rho P} {\bf{w}^H}{{\bf{h}}_{s\jmath}}{x_s^{{\ell}(1)}} + \sqrt {\left( {1 - \rho } \right)P} {h_{\jmath d}}{x_d^{\ell}} + {n_\jmath^{\ell(1)}}},&{{\Psi_1}},
	\end{array}} \right.
\end{align}
where ${\bf{w}} = \frac{{{{\bf{h}}_{si}}}}{{\left\| {{{\bf{h}}_{si}}} \right\|}}$, is the MRT beamformer at the source and $i$ ($ i  \in \left\{ {1,2,...J} \right\}$) displays the index of the selected relay. In the second phase, the selected relay amplifies the received signal by an amplification factor of $ G_i$. 
 According to Section \ref{System model},  the received signal at the destination is given by  
\begin{align}
{y_d^{\ell}}  =\left\{ {\begin{array}{*{20}{l}}
	{n_d^{\ell}},&{{\Psi_0}},\\\\
	{G_i \sqrt {\rho P} {{\bf{w}}^H}{{\bf{h}}_{si}}{h_{id}}{x_s^{{\ell}(1)}} + G_i {n_i^{\ell}}{h_{id}} + {n_d^{\ell}}},&{{\Psi_1}}.
	\end{array}} \right.
\end{align}
As mentioned, in the second phase all the non-selected untrusted relays except of selected relay hear the broadcasted signal by selected relay.  Therefore, the received signal at the untrusted relay $\jmath$ ($\jmath \ne i$) is 
\begin{align} \label{yj1_s}
&y_\jmath ^{\ell (2)} =\\& \left\{ {\begin{array}{*{20}{l}}
	{\sqrt {\xi P} {h_{s\jmath }}x_s^{\ell (2)} + n_\jmath ^{\ell (2)},}&{{\Psi _0},}\\
	\begin{array}{l}
	\\	
	\end{array}&{}\\
	\begin{array}{l}
	\left( {\sqrt {\rho P} {{\bf{w}}^{\bf{H}}}{{\bf{h}}_{si}}x_s^{\ell (1)} + \sqrt {\left( {1 - \rho } \right)P} {h_{id}}x_d^\ell  + n_i^\ell } \right)  \\ G_i{h_{i\jmath }}+
	\sqrt {\xi P} {h_{s\jmath }}x_s^{\ell (2)} + n_\jmath ^{\ell (2)},
	\end{array}&{{\Psi _1},}
	\end{array}} \right. \nonumber
\end{align}
By employing \eqref{yr1_s2}, the SINR at the selected relay i.e., relay $i$ is $\gamma_i=\gamma_R$ (see equation \eqref{g_r0}) and  the other relays  in the first phase are obtained as
\begin{align}\label{g_1_s2}
&	{\gamma _\jmath^{(1)}} =\left\{ {\begin{array}{*{20}{l}}
	{0},&{{\Psi_0}},\\\\
	{ \frac{{\rho {\gamma _{s \jmath}^{bf}}}}{{\left( {1 - \rho } \right){\gamma _{\jmath d}} + 1}} \mathop  = \limits^{(a)}\frac{{\rho \varsigma_\jmath^{bf} }}{{1 - \rho }}},&{{\Psi_1}},
	\end{array}} \right. 
\end{align}
where  
 ${\gamma _{\jmath d}} = \frac{{P{{\left\| {{h_{\jmath d}}} \right\|}^2}}}{\sigma ^2}$, ${\gamma _{s \jmath}^{bf}}= \frac{{P{{\left\| {{{\frac{{{\bf{h}_{si}^H}}}{{\left\| {{{\bf{h}}_{si}}} \right\|}}}}{{\bf{h}}_{s \jmath}}} \right\|}^2}}}{{{\sigma ^2}}}$,  $\varsigma_\jmath^{bf}= \frac{\gamma _{s \jmath}^{bf}}{\gamma _{\jmath d}}$, and $a$ follows high SINR assumption.
Moreover, by utilization \eqref{yj1_s}, the received SINR at the untrusted relay $\jmath \ne i$  during the second phase is given by
\begin{align}\label{snr_r2}
\gamma _\jmath ^{\ell (2)} = \left\{ {\begin{array}{*{20}{l}}
	{0,}&{{\Psi _0},}\\
	{}&{}\\
	{\frac{{\rho {\gamma _{si}}{\gamma _{i\jmath }}\left( {1 - \xi } \right)}}{{\left( {1 - \xi } \right){\gamma _{i\jmath }}\left( {1 + \left( {1 - \rho } \right){\gamma _{id}}} \right) + \left( {\xi {\gamma _{s\jmath }} + 1} \right)\left( {\rho {\gamma _{si}} + \left( {1 - \rho } \right){\gamma _{id}} + 1} \right)}},}&{{\Psi _1}.}
	\end{array}} \right.
\end{align}

Furthermore, the received  SINR at the destination can be written as follows
\begin{align}\label{g_d1}
	{{\gamma _{{D_i}}} = \left\{ {\begin{array}{*{20}{l}}
			{0,}&{{\Psi _0},}\\
			{}&{}\\
			{\frac{{\rho {\gamma _{si}}{\gamma _{id}}\left( {1 - \xi } \right)}}{{\rho {\gamma _{si}} + \left( {2 - \rho  - \xi } \right){\gamma _{id}} + 1}} \mathop  = \limits^{(a)} \frac{{\rho \varsigma_i {\gamma _{id}}\left( {1 - \xi } \right)}}{{\rho \varsigma_i  + 2 - \rho  - \xi }},}&{{\Psi _1},}
			\end{array}} \right.}
\end{align}
Supposing the $i^{th}$ relay is selected, therefore, the instantaneous secrecy rate can be obtained as follows,:
\begin{align}\label{RS}
R_{\sec }^i = \frac{1}{2}{\left[ {\log \left( {1 + {\gamma _{{D_i}}}} \right) - \log \left( {1 + {\Gamma _E}} \right)} \right]^ + },
\end{align}
where $\Gamma _E$ is maximum of  information leaked to the untrusted relays in two phases. As the untrusted relay is non-colluding, $\Gamma _E$ can be expressed as
\begin{align}\label{Gamma_E}
{\Gamma_E} = \mathop {\max }\limits_{i,\jmath  \in \left\{ {1,2,...J} \right\}|i \ne \jmath } \left\{ {\gamma _\jmath ^{\ell (1)},{\gamma _i},\gamma _\jmath ^{\ell (2)}} \right\}.
\end{align}

\textbf{Proposition 1:}
\textit{The amount of the information leaked to the selected relay in the first phase is more than the other relays in the both phases.}

\textit{Proof:}
 It is clear ${\gamma _\jmath ^{\ell (1)} = {\gamma _i} = \gamma _\jmath ^{\ell (2)}}$, when $\Psi_0$ happens. when $\Psi_1$ happens after some manipulation we have

\begin{align}\label{gl2}
\begin{array}{l}
{\gamma _\jmath ^{\ell (2)}}=\frac{{\rho {\gamma _{si}}{\gamma _{i\jmath }}\left( {1 - \xi } \right)}}{{\left( {1 - \xi } \right){\gamma _{i\jmath }}\left( {1 + \left( {1 - \rho } \right){\gamma _{id}}} \right) + \left( {\xi {\gamma _{s\jmath }} + 1} \right)\left( {\rho {\gamma _{si}} + \left( {1 - \rho } \right){\gamma _{id}} + 1} \right)}} < \\
\frac{{\rho {\gamma _{si}}{\gamma _{i\jmath }}\left( {1 - \xi } \right)}}{{\left( {1 - \xi } \right){\gamma _{i\jmath }}\left( {1 + \left( {1 - \rho } \right){\gamma _{id}}} \right)}} = \frac{{\rho {\gamma _{si}}}}{{\left( {1 + \left( {1 - \rho } \right){\gamma _{id}}} \right)}}=\gamma_i,
\end{array}
\end{align} 
 because ${\left( {\xi {\gamma _{s\jmath }} + 1} \right)\left( {\rho {\gamma _{si}} + \left( {1 - \rho } \right){\gamma _{id}} + 1} \right)}$ is positive. Furthermore,  as there is 
large-scale multiple antennas at the source, by benefiting  Cauchy-Schwarz inequality  the upper bound of $\gamma _\jmath ^{\ell (1)}$ can be written as
\begin{align}\label{gl1}
\gamma _\jmath ^{\ell (1)} = \frac{{\rho \gamma _{i\jmath }^{bf}}}{{\left( {1 - \rho } \right){\gamma _{\jmath d}} + 1}} < \frac{{\rho {\gamma _{si}}}}{{\left( {1 + \left( {1 - \rho } \right){\gamma _{id}}} \right)}} = {\gamma _i}.
\end{align} 

Based on \eqref{gl1}, \eqref{gl2}, and \eqref{Gamma_E} the information leakage is equivalent to ${\Gamma_E} = \mathop {\max }\limits_{i,\jmath  \in \left\{ {1,2,...J} \right\}|i \ne \jmath } \left\{ {\gamma _\jmath ^{\ell (1)},{\gamma _i},\gamma _\jmath ^{\ell (2)}} \right\}=\gamma _i$. Hence, the instantaneous secrecy rate can be simplify as follows
\begin{align}\label{RS_simp}
R_{\sec }^i = \frac{1}{2}{\left[ {\log \left( {1 + {\gamma _{{D_i}}}} \right) - \log \left( {1 + {\gamma _i}} \right)} \right]^ + }.
\end{align}
According to Subsection \ref{Physical layer security}, the operator ${\left[ . \right]^ + }$ is ignorable.
\subsection{Relay Selection Criterion}

In this subsection, we study two cases for the relay selection: 1) Optimal relay selection, 2) suboptimal relay selection with the assumption  large-scale multiple antennas (LSMA) at the source.
\subsubsection{Optimal Relay Selection}
With the aim of maximizing the secrecy rate, we select a relay which leads to increase secrecy rate. The optimization problem can be formulated as follows
\begin{subequations}\label{OPT_reformulated_RS}
	\begin{align}
	&\max_{\rho, \xi, t_0, \textbf{T}}\; \mathop {\max }\limits_{i  \in \left\{ {1,2,...J} \right\} } \frac{pr_t}{2}\left[ {\log \left( { \frac{{\rho  \varsigma_i {\gamma _{id}}\left( {1 - \xi } \right)}}{{\rho  \varsigma_i  + 2 - \rho  - \xi }}} \right)} \right.\\&
	{\left. { - \log \left( {1 + \frac{{\rho  \varsigma_i }}{{1 - \rho }}} \right)} \right] }, \nonumber
	\\&
	\hspace{.01cm}\text{s.t.}:\hspace{.18cm} \hspace{.008cm} \eqref{rho_limited}, \eqref{xi_limited},  \eqref{cp01}, \eqref{cp11}
	\\& \hspace{.35cm}  \ln \left( {\frac{{\xi {\mu _{sw}}}}{{\left( {1 - \xi } \right){\mu _{iw}}}}} \right) \times \xi {\mu _{sw}} - {t_i}\ln \left( \varepsilon  \right) \le 0, \, \forall i \in \left\{ {1,2,...J} \right\} \label{cp02_RS}
	\\& \hspace{.35cm} \left( {1 - \xi } \right){\mu _{iw}} - \xi {\mu _{sw}} \le {t_i}, \, \forall i  \in \left\{ {1,2,...J} \right\} \label{cp12_RS}
\end{align}
\end{subequations}
where $ \varsigma_i=\frac{{{\gamma _{si}}}}{{{\gamma _{id}}}}$ and $\textbf{T}= \left\{ {t_1,t_2,...t_J} \right\}$. In order to solve the optimization problem \eqref{OPT_reformulated_RS} we define a slack variable $\tilde t$ as follows
\begin{align}
\mathop {\max }\limits_{\jmath \in \left\{ {1,2,...J} \right\}} \left[ {\log \left( {\frac{{\rho {\varsigma _i}{\gamma _{id}}\left( {1 - \xi } \right)}}{{\rho {\varsigma _i} + 2 - \rho  - \xi }}} \right)} \right.\left. { - \log \left( {1 + \frac{{\rho {\varsigma _i}}}{{1 - \rho }}} \right)} \right]=\tilde t,
\end{align}

By employing the epigraph method the optimization problem \eqref{OPT_reformulated_RS} can be rewritten as follows

\begin{subequations}\label{OPT_reformulated_RS_epg}
	\begin{align}
	&\max_{\rho, \xi, t_0, \textbf{T}, \tilde t}\; \frac{{p{r_t}}}{2} \tilde t, \nonumber
	\\&
	\hspace{-.5cm}\text{s.t.}:\hspace{.18cm} \hspace{.008cm} \eqref{rho_limited}, \eqref{xi_limited},  \eqref{cp01}, \eqref{cp11}, \eqref{cp02_RS}, \eqref{cp12_RS}, 
	\\& \hspace{.35cm}  \log \left( {\frac{{\rho {\varsigma _i}{\gamma _{id}}\left( {1 - \xi } \right)}}{{\rho {\varsigma _i} + 2 - \rho  - \xi }}} \right) - \log \left( {1 + \frac{{\rho {\varsigma _i}}}{{1 - \rho }}} \right) \label{cp02_RS_epi} \\& \hspace{.25cm}  \le \tilde t, \forall \jmath  \in \left\{ {1,2,...J} \right\}. \nonumber
	\end{align}
In order to solve \eqref{OPT_reformulated_RS_epg},  we use the DC method similar to \ref{pa_subsection}, but CVX solver illustrates this  optimization problem is infeasible.
\end{subequations}
\subsubsection{Suboptimal Relay Selection}
We propose a suboptimal relay selection with the assumption of an LSMA at the source. The instantaneous secrecy rate is
\begin{align}\label{SR_aid}
R_{\sec }^i=\frac{1}{2}\log \left( {\frac{{1 + {\gamma _{{D_i}}}}}{{1 + {\gamma _i}}}} \right)
\end{align}
By substituting \eqref{g_d1} and \eqref{g_1_s2} into \eqref{SR_aid}, we have
$\log \left( {\frac{{1 + {\gamma _{{D_i}}}}}{{1 + {\gamma _i}}}} \right) = \log \left( {\frac{{\left( {{\gamma _{id}}\rho {\varsigma _i}\left( {1 - \xi } \right) - \xi  + 2 - \rho  + \rho {\varsigma _i}} \right)\left( {1 - \rho } \right)}}{{\left( { - \xi  + 2 - \rho  + \rho {\varsigma _i}} \right)\left( {\rho {\varsigma _i} - \rho  + 1} \right)}}} \right)$, which is a function of random variables $\gamma_{si}$ and $\gamma_{id}$. Owing to an LSMA at the source and adopting the  law of large numbers, we can approximate $\gamma_{si}$ as $\gamma_{si}\simeq N_s\times \frac{P\mu_{si}}{\sigma^2}$.  With this approximation the secrecy rate is only a function of $\gamma_{id}$ and it is easy to show the secrecy rate is an increasing function with respect to $\gamma_{id}$. Hence, the suboptimal relay selection can be rewritten as follows
\begin{align}\label{Suboptima_RS}
{i^*}=\mathop {\max }\limits_{i  \in \left\{ {1,2,...J} \right\} } \frac{{1 + {\gamma _{{D_i}}}}}{{1 + {\gamma _i}}}\simeq\mathop {\max }\limits_{i  \in \left\{ {1,2,...J} \right\}} {\left| {{h_{id}}} \right|^2}.
\end{align}
As can be seen, the proposed suboptimal relay selection in \eqref{Suboptima_RS} enjoys from the low complexity. 

\section{Multiple Willies Scenario}\label{Multiple Willies Scenario}
In this section, we extend our system model to a practical scenario at which there are $\text{W}$ Willies, $W=\left\{ {{w_1},{w_2},...,{w_W}} \right\}$ in our considered network.
For such a network, we investigate two cases 1) Non Colluding Willies, 2) Colluding Willies. In the following, we explain each of these cases separately. 
\subsection{Non-Colluding Willies}
 In this subsection, we assume the Willies are non-colluding, i.e., each of them separately tries to detect that the data is transmitted or not. Hence, for power allocation, we should first select the Willie with the lowest error detection  (i.e.,  the worst Willie should be selected) and then we follow the same power allocation strategy mentioned in Section \ref{Multi Untrusted Relay Scenario With Relay Selection}.  It  is clear that the expressions \eqref{sum_fa_md_1} and \eqref{sum_fa_md_2} are decreasing functions with respect to $ \frac{d_{dw}}{d_{sw}}$ and $\frac{d_{sw}}{d_{rw}}$, respectively. Therefore, the worst Willie in first and second phases are
$
{w^{*(1)}}=\mathop {\max }\limits_{w_i  \in W}  \frac{d_{dw_{i}}}{d_{sw_{i}}}
$
and 
$
{w^{*(2)}}=\mathop {\max }\limits_{w_i  \in W} \frac{d_{sw_{i}}}{d_{rw_{i}}}
$, respectively.
\subsection{Colluding Willies}
In this subsection, we assume the Willies are colluding, i.e., each of them delivers its received signal energy to a Fusion Center (FC) to decide on the presence or absence of the data transmission. Therefore, the received signal energy at the FC in the first and second phases are  $Y_{FC}^{(1)} = \sum\limits_{{w_i} = 1}^W {\sum\limits_{\ell  = 1}^n {{{\left| {y_{{w_i}}^{\ell (1)}} \right|}^2}} } $ and $Y_{FC}^{(2)}  = \sum\limits_{{w_i} = 1}^W {\sum\limits_{\ell  = 1}^n {{{\left| {y_{{w_i}}^{\ell (2)}} \right|}^2}} } $, respectively. 
Similar to Section \ref{Power Allocation for  Secrecy Rate Maximization With Covert Communication Requirement} and this fact that the summation of $n$ independent exponential random variables $X_i$ with different parameters $\lambda_i$ have  the probability density function of $f_{X_1+X_2+\text{...}+X_n}(x){\rm{ = }}\left[ {\mathop \Pi \limits_{i = 1}^n {\lambda _i}\sum\limits_{j = 1}^n {\frac{{{e^{ - {\lambda _j}x}}}}{{\mathop \Pi \limits_{\scriptstyle k = 1\hfill\atop
\scriptstyle k \ne j\hfill}^n \left( {{\lambda _k} - {\lambda _j}} \right)}}} } \right]\,\,\,\,\,\,\,x > 0 $, the probability of false alarm and miss detection are respectively, given by
\begin{align}
\label{pfa_mW}\mathbb{P}_{FA}^{(1)} = \left\{ {\begin{array}{*{20}{l}}
	{\mathop \Pi \limits_{i = 1}^W {\hat \tau _i^{(1)}}\sum\limits_{j = 1}^W {\frac{{{e^{ - {\hat \tau _j^{(1)}}\left( {{\vartheta ^{(1)}} - W\sigma _w^2} \right)}}}}{{\mathop \Pi \limits_{\scriptstyle k = 1\hfill\atop
						\scriptstyle k \ne j\hfill}^W \left( {{\hat \tau _k^{(1)}} - {\hat \tau _j^{(1)}}} \right)}}} {\mkern 1mu} {\mkern 1mu} {\mkern 1mu} ,}&{{\vartheta ^{(1)}} - W\sigma _w^2 > 0,}\\
	{}&{}\\
	{1,}&{{\vartheta ^{(1)}} - W\sigma _w^2 \le 0,}
	\end{array}} \right.
\\\label{pmd_mW}
\mathbb{P}_{MD}^{(1)} = \left\{ {\begin{array}{*{20}{l}}
	{1 - \mathop \Pi \limits_{i = 1}^{2W} {\tilde \tau _i^{(1)}}\sum\limits_{j = 1}^{2W} {\frac{{{e^{ - {\tilde \tau _j^{(1)}}\left( {{\vartheta ^{(1)}} - W\sigma _w^2} \right)}}}}{{\mathop \Pi \limits_{\scriptstyle k = 1\hfill\atop
						\scriptstyle k \ne j\hfill}^{2W} \left( {{\tilde \tau _k^{(1)}} - {\tilde \tau _j^{(1)}}} \right)}}} }&{{\vartheta ^{(2)}} - W\sigma _w^2 > 0,}\\
	{\begin{array}{*{20}{l}}
		{}
		\end{array}}&{}\\
	0&{{\vartheta ^{(2)}} - W\sigma _w^2 \le0,}
	\end{array}} \right.
\end{align}
where $\hat \tau _i^{(1)} = \left({\left( {1 - \rho } \right)P{\mu _{d{w_i}}}}\right)^{-1}, \,\, \,i = 1,...,W$,  $\tilde \tau _i^{(1)} = \hat \tau _i^{(1)},\,\,\,i = 1,...,W$, and $\tilde \tau _i^{(1)} = \left({\rho P{\mu _{s{w_i}}}} \right)^{-1},\,\,i = W + 1,...2W$. Likewise, one can obtain the probability of false alarm and miss detection in the second phase. By substituting the calculated probability of false alarm and miss detection into \eqref{OPT}, the mentioned optimization problem is intractable and its solution is very complex, especially for  large number of Willies in the network.  To tackle this issue, we use the central limit theorem (CLT) to calculate the probability of false alarm and miss detection in the both phases. In order to simplify mathematics, we assume  the Willi's channels are independent and identically distributed (iid), i.e.,  $\mu_{sw_i}=\mu_{sw_j}$, $\mu_{rw_i}=\mu_{rw_j}$, $\mu_{dw_i}=\mu_{dw_j}$, $\forall w_i\,\& \,w_j \in W$, \cite{L.Wang}, \cite{Krikidis}.
 Using of the CLT leads to the following equations 
\begin{align}
&\mathbb{P}_{FA}^{(1)} =Q\left( {\frac{{{\vartheta ^{(1)}} - {\mu _{fa}^{(1)}}}}{{{\sigma _{fa}^{(1)}}}}} \right), \,\,\,
\mathbb{P}_{MD}^{(1)} =1 - Q\left( {\frac{{{\vartheta ^{(1)}}- {\mu _{md}^{(1)}}}}{{{\sigma _{md}^{(1)}}}}} \right),\\
&\mathbb{P}_{FA}^{(2)} =Q\left( {\frac{{{\vartheta ^{(2)}} - {\mu _{fa}^{(2)}}}}{{{\sigma _{fa}^{(2)}}}}} \right), \,\,\,
\mathbb{P}_{MD}^{(2)} =1 - Q\left( {\frac{{{\vartheta ^{(2)}}- {\mu _{md}^{(2)}}}}{{{\sigma _{md}^{(2)}}}}} \right),
\end{align}
   \begin{figure*}[t]    	
	\begin{align}
	\label{top1} \vartheta _{opt1}^{(1)} =& \frac{{ - \mu _{fa}^{(1)}\sigma _{md}^{(1)2} + \mu _{md}^{(1)}\sigma _{fa}^{(1)2}}}{{\sigma _{fa}^{(1)2} - \sigma _{md}^{(1)2}}}\nonumber +\\& \frac{{\sqrt {2\ln \left( {\frac{{\sigma _{fa}^{(1)}}}{{\sigma _{md}^{(1)}}}} \right)\sigma _{fa}^{(1)4}\sigma _{md}^{(1)2} - 2\ln \left( {\frac{{\sigma _{fa}^{(1)}}}{{\sigma _{md}^{(1)}}}} \right)\sigma _{fa}^{(1)2}\sigma _{md}^{(1)4} + \mu _{fa}^{(1)2}\sigma _{fa}^{(1)2}\sigma _{md}^{(1)2} - 2\mu _{fa}^{(1)}\mu _{md}^{(1)}\sigma _{fa}^{(1)2}\sigma _{md}^{(1)2} + \mu _{md}^{(1)2}\sigma _{fa}^{(1)2}\sigma _{md}^{(1)2}} }}{{\sigma _{fa}^{(1)2} - \sigma _{md}^{(1)2}}}\\\label{top2}
	\vartheta _{opt2}^{(1)} = &- \frac{{\mu _{fa}^{(1)}\sigma _{md}^{(1)2} - \mu _{md}^{(1)}\sigma _{fa}^{(1)2}}}{{\sigma _{fa}^{(1)2} - \sigma _{md}^{(1)2}}}- \nonumber\\& \frac{{\sqrt {2\ln \left( {\frac{{\sigma _{fa}^{(1)2}}}{{\sigma _{md}^{(1)}}}} \right)\sigma _{fa}^{(1)4}\sigma _{md}^{(1)2} - 2\ln \left( {\frac{{\sigma _{fa}^{(1)}}}{{\sigma _{md}^{(1)}}}} \right)\sigma _{fa}^{(1)2}\sigma _{md}^{(1)4} + \mu _{fa}^{(1)2}\sigma _{fa}^{(1)2}\sigma _{md}^{(1)2} - 2\mu _{fa}^{(1)}\mu _{md}^{(1)}\sigma _{fa}^{(1)2}\sigma _{md}^{(1)2} + \mu _{md}^{(1)2}\sigma _{fa}^{(1)2}\sigma _{md}^{(1)2}} }}{{\sigma _{fa}^{(1)2} - \sigma _{md}^{(1)2}}}
	\end{align}
	\hrule
\end{figure*}
where ${\mu _{fa}^{(1)}} = W\left( {1 - \rho } \right)P{\mu _{dw}}+W\sigma _w^2$, ${\sigma _{fa}^{(1)}} = \sqrt W \left( {1 - \rho } \right)P{\mu _{dw}}$, ${\mu _{md}^{(1)}} =  W\rho P{\mu _{sw}} + W\left( {1 - \rho } \right)P{\mu _{dw}}+W\sigma _w^2$,  ${\sigma _{md}^{(1)}} = \sqrt {W{\rho ^2}{P^2}\mu _{sw}^2 + W{{\left( {1 - \rho } \right)}^2}{P^2}\mu _{dw}^2} $,
 $\mu _{fa}^{(2)} = W\xi P{\mu _{sw}} + W\sigma _w^2$, $\sigma _{fa}^{(2)} = \sqrt W \xi P{\mu _{sw}}$, $\mu _{md}^{(2)} = W\left( {1 - \xi } \right)P{\mu _{rw}} + W\xi P{\mu _{sw}} + W\sigma _w^2$,  $\sigma _{md}^{(2)} = \sqrt {W{{\left( {1 - \xi } \right)}^2}{P^2}\mu _{rw}^2 + W{\xi ^2}{P^2}\mu _{sw}^2} $, and $Q(x)$ is Q-function. In the following, we first find the optimal threshold for Willie and then formulate our optimization problem.

\subsubsection{Optimal Threshold for Willie}
In order to obtain optimal  $\vartheta^{(1)}$, we derivative from ${\mathbb{P}^{(1)}_{FA}} + {\mathbb{P}^{(1)}_{MD}}$ with respect to $\vartheta^{(1)}$.
After calculation 
$\frac{{d\left( {\mathbb{P}^{(1)}_{FA}} + {\mathbb{P}^{(1)}_{MD}} \right)}}{{d\vartheta^{(1)}}}=0$
and some mathematical manipulations, we obtain \eqref{top1} and \eqref{top2} at the top of the this page.  Note that since ${\mu _{md}^{(1)}}\ge {\mu _{fa}^{(1)}}$ and ${\sigma _{md}^{(1)}}\ge {\sigma _{fa}^{(1)}}$, it is clear to show that $\vartheta _{opt1}^{(1)} \le 0$ and $\vartheta _{opt2}^{(1)} \ge 0$ and hence, $\vartheta _{opt2}^{(1)}$ is acceptable. Likewise,  $\vartheta _{opt1}^{(2)}$ and $\vartheta _{opt2}^{(2)}$ are obtainable.
\subsubsection{Optimization problem}
Similar to Section \ref{Power Allocation for  Secrecy Rate Maximization With Covert Communication Requirement}, the optimization problem  can be written as follows
\begin{subequations}\label{opt_MW}
\begin{align}\label{op_mW}
&\max_{\rho, \xi}\; \frac{pr_t}{2} \left( {\Sigma \left( {\rho ,\xi } \right) - \tilde \Omega \left( {\rho ,\xi } \right)} \right)
\\&
\hspace{-.5cm}\text{s.t.}:\hspace{.18cm} \hspace{.008cm}\eqref{rho_limited}, \eqref{xi_limited},
\\&
 \hspace{.35cm} \label{co1}1 - Q\left( {\frac{{\vartheta _{opt2}^{(1)}- {\mu _{md}^{(1)}}}}{{{\sigma _{md}^{(1)}}}}} \right)+ Q\left( {\frac{{\vartheta _{opt2}^{(1)} - {\mu _{fa}^{(1)}}}}{{{\sigma _{fa}^{(1)}}}}} \right)\ge 1- \varepsilon,
 \\&
 \hspace{.35cm} \label{co2} 1 - Q\left( {\frac{{\vartheta _{opt2}^{(2)}- {\mu _{md}^{(2)}}}}{{{\sigma _{md}^{(2)}}}}} \right)+ Q\left( {\frac{{\vartheta _{opt2}^{(2)} - {\mu _{fa}^{(2)}}}}{{{\sigma _{fa}^{(2)}}}}} \right)\ge 1- \varepsilon.
\end{align}
\end{subequations}
The left side of \eqref{co1} is a decreasing function of $\rho$. Therefore, the  constraint \eqref{co1} is equivalent to $\rho \le \rho_{ub}$. Furthermore, the left side of \eqref{co2} is an increasing function  of $\xi$, therefore, the constraint \eqref{co1} is equivalent to $\xi \ge \xi_{lb}$. Conclusively, the optimization problem \eqref{op_mW} is equivalent to the following convex optimization problem
\begin{subequations}\label{opt_MW1}
	\begin{align}
	&\max_{\rho, \xi}\; \frac{pr_t}{2} \left( {\Sigma \left( {\rho ,\xi } \right) - \tilde \Omega \left( {\rho ,\xi } \right)} \right)
	\\&
	\hspace{-.5cm}\text{s.t.}:\hspace{.18cm} \hspace{.008cm} 0\le \rho \le \rho_{ub},
	\\&
	\hspace{.35cm}  \xi_{lb} \le \xi \le 1.
	\end{align}
\end{subequations}
To solve the convex optimization problem \eqref{opt_MW1}, we use the available softwares such as CVX solver.
\section{ Secure Null Space Beamforming of Direct Transmission Scheme}\label{DT}
In the direct transmission scheme, we assume there is the direct link between the source and the destination and hence, the transmission only is performed in one phase. When the source transmits its signal with $N_s$ antenna, the relays are treated as pure eavesdroppers, i.e., they only listen to the signal. Furthermore, Willies try to detect  whether the source has sent a signal to the destination or not, in the transmission phase. In order to deceive the untrusted relays and Willies, the source transmits information and jamming signal simultaneously, and it employs null space beamforming of jamming signal over the destination's channel, to increase the secrecy rate.  The detailed formulations of this secure transmission scheme is provided in Appendix A.

\section{Numerical Results}\label{Numerical Results}
In this section, numerical results are presented to evaluate the performance of joint covert communication and PLS  in one-way relay network in presence of one source and  one destination, one untrusted relay and one Willie. The used parameters in our simulations are: $\sigma_w^2=\sigma^2=-50$ dBW,  the path loss exponent $\alpha=4$, probability  of transmission in each time slot ${pr_t}=0.5$, ‌the channels are assumed complex Gaussian random variables and the maximum transmission power in each time slot $P=10$ dBW.
\textcolor{black}{  Without loss of generality, we assume the source, the destination, the untrusted relay, and  Willie are  located at $(-5,0)$, $(5,0)$, $(0,0)$, and $(0,-5)$.}

\begin{figure}[t]  	
  	\begin{center}
  		\includegraphics[width=3.8in,height=3in]{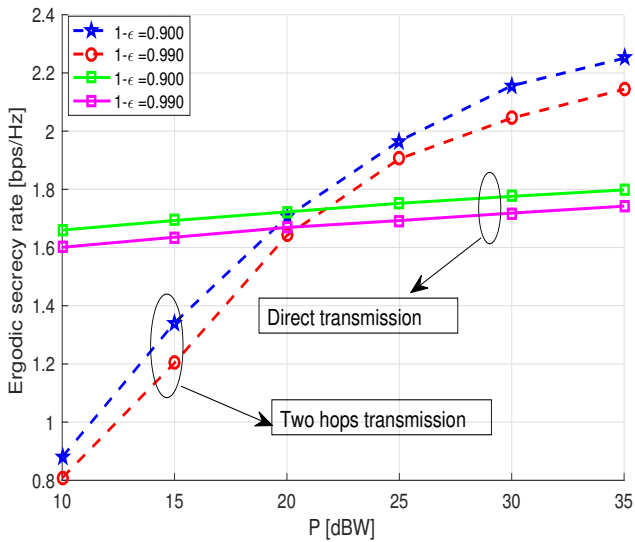}
  		\caption{Ergodic secrecy rate versus maximum transmit power for single relay and Willie, $N_s=16$.}
  		\label{SR_P}
  	\end{center}
  \end{figure}

 Fig. \ref{SR_P} shows the ergodic secrecy rate versus transmission power for single relay and single Willie case. 
As can be seen, the ergodic secrecy rate is an increasing function with respect to the transmit power. The reason is that in the studied scenarios (the direct transmission and the two hops scenarios), the injected jamming signal only degrades the received information signal at the illegitimate nodes and has not impact on the received  information signal at the destination because of employing null space beamforming technique in the direct transmission scenario and interference cancellation in the two hops scenario.
Moreover, the the ergodic secrecy rate of the direct transmission scheme surpass two hops scheme in low transmit power and opposite behavior is observed in high transmit power. This is because by equipping with multiple antenna and employing an MRT beamformer at the source, the information leakage in illegitimate nodes is negligible and the received SNR at the destination is considerable. 

\begin{figure}[t]  	
	\begin{center}
		\includegraphics[width=3.8in,height=3in]{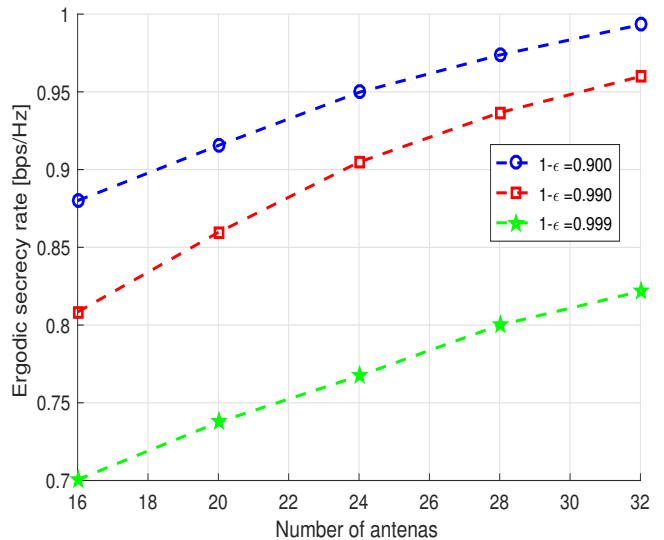}
		\caption{ٍErgodic secrecy rate versus numbers of antenna at source, $P=10$ dBW. }
		\label{SR_Ns}
	\end{center}
\end{figure}

In Fig. \ref{SR_Ns}, the ergodic secrecy rate versus number of antenna at source is shown.
   This figure evaluates effect of lower bound of detection error probability at Willie i.e., $1-\epsilon$. As seen in this figure, guaranty of 99.9\% error in detection of  Willie with respect to  guaranty of 99\% error, decreases the ergodic secrecy rate \%14.35. The reason is that by increasing lower bound of detection error probability, in the first phase,  the source has to decrease the power of information signal  and destination has to  increase the power of jamming signal, furthermore, in the second phase the relay has to decrease the power of information signal  and the source has to  increase the power of jamming signal, which leads to decrease ergodic secrecy rate.

\begin{figure}[t]  	
	\begin{center}
		\includegraphics[width=3.8in,height=3in]{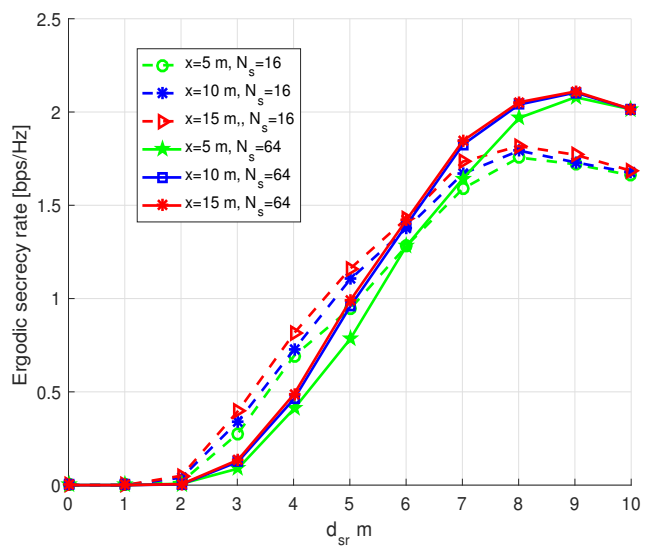}
		\caption{Ergodic secrecy rate versus distance between source and untrusted relay, the number of source antennas $N_s=16$.}
		\label{SR_dsr}
	\end{center}
\end{figure}

\begin{figure}[h]  	
	\begin{center}
		\includegraphics[width=3.2in,height=1.5in]{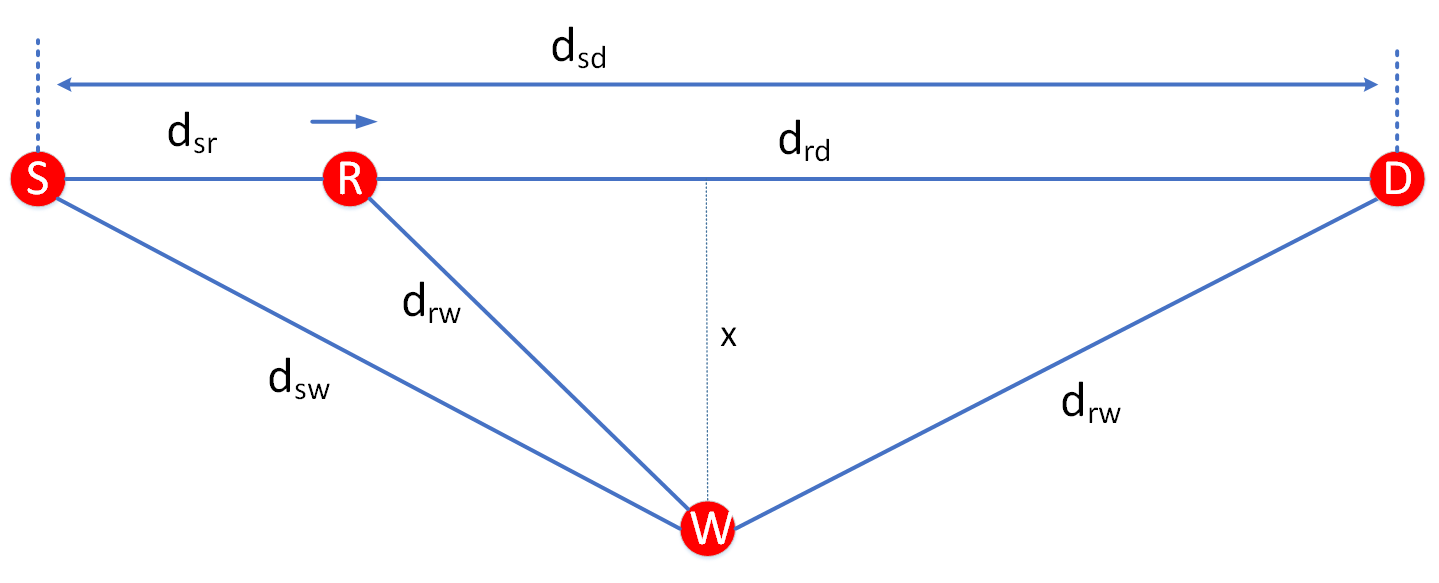}
		\caption{Location schematic of nodes.}
		\label{Mobility}
	\end{center}
\end{figure}

In Fig. \ref{SR_dsr}, we study the impact of the relay's position and  Willie's position on the ergodic secrecy rate. In this figure, we assume the untrusted relay moves from source to destination as depicted in Fig. \ref{Mobility}. The figure  illustrates that for given number of source antennas, there is a optimal relay location which maximizes the ergodic secrecy rate. For example, when $N_s=16$, the optimal distance between source and relay is $d_{sr}=8$ while when $N_s=64$, the optimal distance is $d_{sr}=9$. Fig. \ref{SR_dsr} also illustrates that when the untrusted relay is near to source it is recommended  number of antenna at source  to become lower than when it is far source.

\begin{figure}[t]  	
	\begin{center}
		\includegraphics[width=3.8in,height=3in]{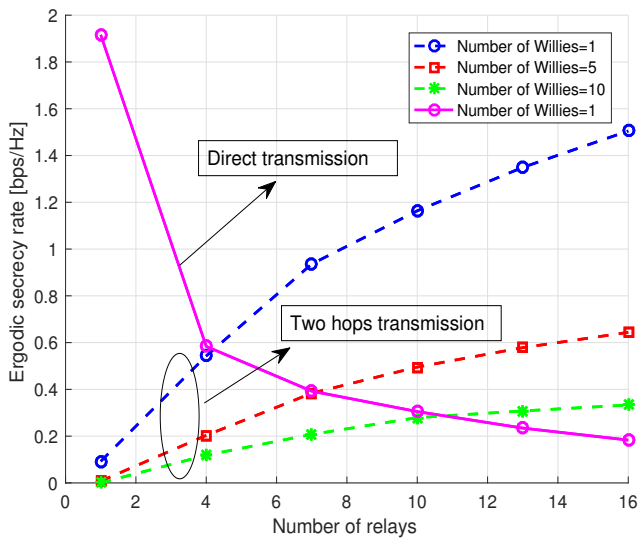}
		\caption{Ergodic secrecy rate versus number of relay, under non-colluding Willies scenario, $N_s=16$, $\epsilon=0.1$.}
		\label{SR_SNR}
	\end{center}
\end{figure}

In Fig. \ref{SR_SNR}, the impact of the number of untrusted relays on the ergodic secrecy rate in our system model is depicted. In this plot, the number of antennas at source is $N_s=16$  \textcolor{black}{and we assume the untrusted relays and Willies are distributed uniformly around of $(0,0)$ and $(0,-5)$, respectively.} We evaluate this scenario for two cases of small, medium, and large number of Willies, $1$, $5$, $10$, respectively. This figure shows that the ergodic secrecy rate is an increasing function of the number of untrusted relays $J$. This means that employing more untrusted relays increases the ergodic secrecy rate of the network. 
	By increasing the number of non-colluding Willies, the ergodic secrecy  rate is decreased, because the source should decrease the information transmit power and increase  the jamming transmit power to satisfy covert requirement which leads to a decrease in the ergodic secrecy rate.
 By increasing the number of Willies from $1$ to $5$, the ergodic secrecy rate decreases about $\%64.15$ while by increasing the number of Willies from $5$ to $10$, it decreases about $\%53.53$.  Furthermore, we have simulated the direct transmission. As can be seen, increasing the number of relays in the two hops scenario increases the ergodic secrecy rate, while the	opposite behavior is observed for the direct transmission scenario.  This is because by increasing $J$ the probability of emerging a stronger relay- destination's channel in the two hops scenario increases  which leads to enhancing the ergodic secrecy rate. While in the direct transmission scenario, by increasing $J$ the probability of emerging a stronger wiretap channel increases which leads to a decrease in the ergodic secrecy rate

\begin{figure}[h]  	
	\begin{center}
		\includegraphics[width=3.8in,height=3in]{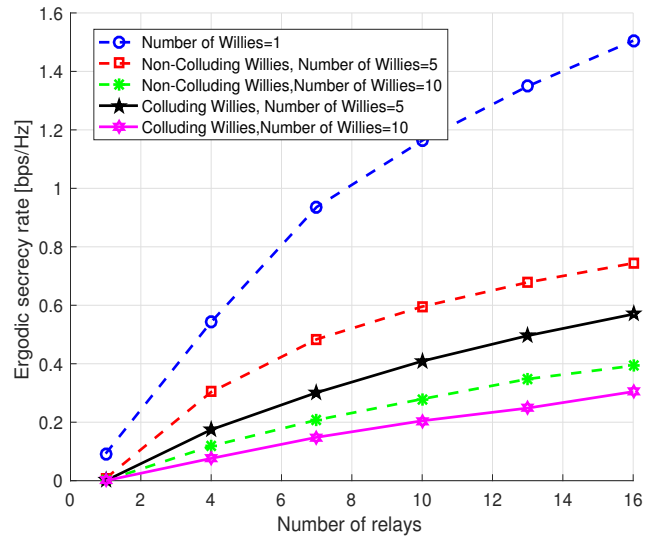}
		\caption{Ergodic secrecy rate versus number of relays under two cases of non-colluding Willies and colluding Willies, $N_s=16$, $\epsilon=0.1$.}
		\label{SR_Colluding} 
	\end{center}
\end{figure}

In Fig. \ref{SR_Colluding}, the impact of the number of untrusted relays on the ergodic secrecy rate for the non-colluding and colluding cases are investigated.
 This figure states that by increasing the number of untrusted relays, the achievable secrecy rate improves for both the non-colluding and colluding Willies. Therefore, this figure shows the priority of the proposed secure transmission scheme compared to the direct transmission scheme ignoring the untrusted relays. Furthermore, as can be seen in this figure, by increasing the number of Willies from $1$ to $5$ for the non-colluding case, the ergodic secrecy rate decreases about $\%55$, while for the colluding case it decreases about $\%71$. Moreover, the network in the presence of $10$ non-colluding Willies compared to $5$ colluding Willies has lower ergodic secrecy rate. As shown, by increasing the number of relays, the secrecy performance of colluding Willies case is lower than the non-colluding case. Because in the colluding Willies case,  Willies combine the received signal energies to decide on the presence or absence of communication which leads to low detection error probability.

 \section{Conclusion}\label{Conclusion}

In this paper, we studied the problem of secure transmission with covert requirement in untrusted relaying networks. Our considered system model consisted of one source, one destination, one untrusted relay, and one Willie. We assumed that the source is equipped with multiple antennas, while the rest are equipped with a single antenna. In our considered system model,  the untrusted relay tried to extract the information signal, while the goal of Willie was to detect  the presence of the information signal transmitted by the source, in the communication phase. To overcome these two attacks, we illustrated that the destination and the source should inject jamming signals to the network in phase I and phase II, respectively. For this system model, we proposed a new power allocation strategy to maximize the instantaneous secrecy rate subject to satisfying the  covert requirements in both of the phases. Since the proposed optimization problem was non-convex, we adopted the Successive Convex Approximation (SCA) approach  to convert it to a convex optimization problem. We next generalized our system model to a more practical communication network where there are multiple untrusted relays and multiple Willies under two cases of non-colluding Willies and colluding Willies.
For the non-colluding case, we proposed a novel optimization problem at which the Willie with the lowest detection error is selected in each phase. Furthermore, for the colluding Willies case,  Willies combine the received signal energies to decide on the presence or absence of communication.
Numerical results revealed that when the untrusted relay is near to the source fewer number of source's antennas is needed compared to the case that the relay is far from the source.  Finally, as a benchmark, we investigated the conventional direct transmission dispensing with the relays to compare with the proposed transmission scheme.

\appendix 
\section*{ Appendix A: Secure Null Space Beamforming of Direct Transmission Scheme}
The received ${\ell}^{th} $ signal at receiver $m$ (relay, destination and Willie) in the direct transmission scheme is given by
\begin{align}
y_m^\ell  = \left\{ {\begin{array}{*{20}{l}}
	{{{\bf{z}}^H}{{\bf{h}}_{sm}}x_{sj}^\ell  + n_m^\ell ,}&{{\Psi _0},}\\
	{}&{}\\
	{\sqrt {\rho P} {{\bf{w}}^{\bf{H}}}{{\bf{h}}_{sm}}x_{si}^\ell  + {{\bf{z}}^H}{{\bf{h}}_{sm}}x_{sj}^\ell  + n_m^\ell ,}&{{\Psi _1},}
	\end{array}} \right.
\end{align}
where weight vector $\bf{w}$ represents the MRT beamformer for data signal at the source and defined as ${\bf{w}} = \frac{{{{\bf{h}}_{sd}}}}{{\left\| {{{\bf{h}}_{sd}}} \right\|}}$. Moreover, $\textbf{z} \in \mathbb{C}^{N_s \times 1}$ is define as the jamming weight vector at the source and ${\left\| {\bf{z}} \right\|} = \sqrt{ (1-\rho)P}$. Note that $x_{sj}^\ell $ and $x_{si}^\ell$ are jamming and information signals, respectively, and the factor $\rho$ is the power allocation between the information bearing signal and the artificial noise emitted by the source.
The received SINR at the destination and the $\jmath$th  relay are respectively, given by
\begin{align}\label{bim_gd}
{{\gamma _{{D}}} = \left\{ {\begin{array}{*{20}{l}}
		{0,}&{{\Psi _0},}\\
		{}&{}\\
		{\frac{{\rho P{{\left\| {{{\bf{w}}^{\bf{H}}}{{\bf{h}}_{sd}}} \right\|}^2}}}{{{{\left\| {{{\bf{z}}^{\bf{H}}}{{\bf{h}}_{sd}}} \right\|}^2} + {\sigma ^2}}},}&{{\Psi _1},}
		\end{array}} \right.}
	\,\,\,
{{\gamma _{{\jmath}}} = \left\{ {\begin{array}{*{20}{l}}
		{0,}&{{\Psi _0},}\\
		{}&{}\\
		{\frac{{\rho P{{\left\| {{{\bf{w}}^{\bf{H}}}{{\bf{h}}_{s\jmath}}} \right\|}^2}}}{{{{\left\| {{{\bf{z}}^{\bf{H}}}{{\bf{h}}_{s\jmath}}} \right\|}^2} + {\sigma ^2}}},}&{{\Psi _1}.}
		\end{array}} \right.}
\end{align}
We assume the relays are non-clouding, hence, the secrecy rate is given by
\begin{align}\label{bim_secrecy rate}
{R_{\sec }} = {\left[ {\log \left( {1 + {\gamma _D}} \right) - \log \left( {\mathop {\max }\limits_{\jmath \in \left\{ {1,2,...J} \right\}} \left( {1 + {\gamma _\jmath}} \right)} \right)} \right]^ + }.
\end{align}
We formulate the optimization problem which its aim is to maximize the worst-case secrecy rate as:

\begin{subequations}\label{bim_OPT_reformulated0}
	\begin{align}
	&\max_{\rho,\textbf{z}, t_0}\; pr_t \left[{{\log \left( {1 + {\gamma _D}} \right) - \log \left( {\mathop {\max }\limits_{\jmath \in \left\{ {1,2,...J} \right\}} \left( {1 + {\gamma _\jmath}} \right)} \right)}}\right]
	\\&
	\hspace{-.5cm}\text{s.t.}:\hspace{.18cm} \hspace{.008cm}0\le \rho\le1 \label{bim_c1},
	\\&‌\hspace{.35cm}\ln \left( {\frac{{{{\left\| {\bf{z}} \right\|}^2}}}{{\rho P{}}}} \right) \times {\left\| {\bf{z}} \right\|^2}{} - {t_0}\ln \left( \varepsilon  \right) \le 0,\label{bim_c2}
	\\&‌\hspace{.35cm} \rho P {} - {{\left\| {\bf{z}} \right\|}^2} \le {t_0}, \label{bim_c3}
	\\&‌\hspace{.35cm} \textbf{z}^Hh_{sd}=0,\label{bim_c4}
		\\&‌\hspace{.35cm} {\left\| {\bf{z}} \right\|^2} = (1-\rho)P. \label{bim_c5}
	\end{align}
\end{subequations}
We define the slack variable $\nu$ as $\frac{1}{\nu } = \mathop {\max }\limits_{\jmath  \in \left\{ {1,2,...J} \right\}} \left( {1 + {\gamma _\jmath }} \right)$, hence, the optimization problem  \eqref{bim_OPT_reformulated0} can be rewritten as follow

\begin{subequations}\label{bim_OPT_reformulated1}
	\begin{align}
	&\max_{\rho,\textbf{z}, t_0}\; pr_t \left[{{\log \left( {1 + {\frac{{\rho P{{\left\| {{{\bf{w}}^{\bf{H}}}{{\bf{h}}_{sd}}} \right\|}^2}}}{\sigma ^2}}} \right) + \log \left( \nu \right)}}\right]
	\\&
	\hspace{-.5cm}\text{s.t.}:\hspace{.18cm} \hspace{.008cm}\eqref{bim_c1}, \eqref{bim_c4}, \nonumber
	\\&‌\hspace{.35cm}\ln \left( {\frac{{{(1-\rho)P}}}{{\rho P{}}}} \right) \times {(1-\rho)P}{} - {t_0}\ln \left( \varepsilon  \right) \le 0,\label{bim_c2_1}
	\\&‌\hspace{.35cm} \rho P {} - {(1-\rho)P} \le {t_0}, \label{bim_c3_1}
	\\&‌\hspace{.35cm} \rho P{\left\| {{{\bf{w}}^{\bf{H}}}{{\bf{h}}_{sj}}} \right\|^2} + {\left\| {{{\bf{z}}^{\bf{H}}}{{\bf{h}}_{s\jmath}}} \right\|^2}  + {\sigma ^2}\le  \label{bim_tylor} \\& \frac{{{{\left\| {{{\bf{z}}^{\bf{H}}}{{\bf{h}}_{s\jmath}}} \right\|}^2}}+\sigma^2}{{\nu}}, \, \forall \jmath \in \left\{ {1,2,...J} \right\}. \nonumber
	\end{align}
\end{subequations}
The Constraint \eqref{bim_tylor} is non-convex because of term $\frac{{{{\left\| {{{\bf{z}}^{\bf{H}}}{{\bf{h}}_{s\jmath}}} \right\|}^2}}+\sigma^2}{{\nu}}$. In order to convert the optimization problem to convex one, we replace this
quadratic-over-linear function with their corresponding first order expansions. The first-order Taylor expansion of the right side of Constraint \eqref{bim_tylor}  at a point of $(\bf{\tilde z}, \tilde \nu)$ is $\frac{2}{{\tilde \nu }} - \frac{\nu }{{{{\tilde \nu }^2}}} + \frac{{2{\mathop{\rm Re}\nolimits} \left\{ {{{{\bf{\tilde z}}}^H}{H_{sj}}{\bf{z}}} \right\}}}{{\tilde \nu }} - \frac{{{{{\bf{\tilde z}}}^H}{H_{sj}}{\bf{z}}}}{{\tilde \nu }}\nu $, \cite{Lv2015}. Finally, we solve the following convex optimization problem
\begin{subequations}
	\begin{align}
	&\max_{\rho,\textbf{z}, t_0}\; pr_t \left[{{\log \left( {1 + {\frac{{\rho P{{\left\| {{{\bf{w}}^{\bf{H}}}{{\bf{h}}_{sd}}} \right\|}^2}}}{\sigma ^2}}} \right) + \log \left( \nu \right)}}\right] 
	\\&
	\hspace{-.1cm}\text{s.t.}:\hspace{.18cm} \hspace{.008cm}\eqref{bim_c1}, \eqref{bim_c4},  \eqref{bim_c2_1}, \eqref{bim_c3_1}, \nonumber
	\\&‌\hspace{.35cm} \rho P{\left\| {{{\bf{w}}^{\bf{H}}}{{\bf{h}}_{sj}}} \right\|^2} + {\left\| {{{\bf{z}}^{\bf{H}}}{{\bf{h}}_{s\jmath}}} \right\|^2}  + {\sigma ^2}\le \\& \hspace{.35cm}  \frac{2}{{\tilde \nu }} - \frac{\nu }{{{{\tilde \nu }^2}}} + \frac{{2{\mathop{\rm Re}\nolimits} \left\{ {{{{\bf{\tilde z}}}^H}{H_{sj}}{\bf{z}}} \right\}}}{{\tilde \nu }} - \frac{{{{{\bf{\tilde z}}}^H}{H_{sj}}{\bf{z}}}}{{\tilde \nu }}\nu, \, \forall \jmath \in \left\{ {1,2,...J} \right\}.\nonumber
	\end{align}
\end{subequations}

\end{document}